\pdfoutput=1
\documentclass[suppldata]{interact}

\usepackage{epstopdf}
\usepackage[caption=false]{subfig}

\usepackage{url}

\usepackage[numbers,sort&compress]{natbib}
\bibpunct[, ]{[}{]}{,}{n}{,}{,}

\usepackage{booktabs}
\usepackage{multirow}

\usepackage{graphicx}

\usepackage{booktabs, longtable, array, xcolor, multirow}
\usepackage[normalem]{ulem} 
\usepackage[table]{xcolor}

\definecolor{rowgray}{gray}{0.95}

\definecolor{rowgray}{gray}{0.93}

\theoremstyle{plain}

\theoremstyle{definition}

\theoremstyle{remark}

\usepackage{hyperref}
\hypersetup{
    colorlinks=true,
    allcolors=black
}

\begin{document}

\articletype{RESEARCH ARTICLE}

\title{Who Trusts AI with Their Emotions? Trust Formation and Sociodemographic Variation in LLM Use for Emotional Support}


\author{
\name{Natalia Amat-Lefort\textsuperscript{a}\thanks{Corresponding: Natalia Amat-Lefort. Email: n.amat.lefort@liacs.leidenuniv.nl}, Mert Yazan\textsuperscript{a,b}, Amanda Cercas Curry\textsuperscript{c}, and Flor Miriam Plaza-del-Arco\textsuperscript{a}}
\affil{\textsuperscript{a}Leiden Institute of Advanced Computer Science, Leiden University. Leiden (Netherlands); \textsuperscript{b}Hogeschool van Amsterdam. Amsterdam (Netherlands); \textsuperscript{c}Independent Researcher}
}

\maketitle

\begin{abstract}
Trust in AI for emotional support is not universal; it is shaped by who users are, where they come from, and what they value. Yet research in this area lacks validated psychometric instruments for assessing user perceptions in affective AI contexts and large-scale evidence on how trust formation varies across user segments. 
To address these gaps, we develop and validate a seven-construct psychometric scale, test a Structural Equation Model (SEM) linking system attributes to Trust and Perceived Benefits as mediators of Actual System Use, and conduct a Multi-Group Analysis (MGA) across five sociodemographic dimensions (gender, age, education, socioeconomic status, cross-national region), drawing on 1,343 active users from seven countries. 
We find that users experience empathy and anthropomorphism as a unified ``Humanlikeness'' construct, and that Privacy, Personalization, and Humanlikeness drive Trust while Perceived Bias degrades it. 
Notably, adoption logic diverges across groups: Privacy shapes women's trust more than men's, Anglosphere (UK, USA) users respond more positively to Humanlikeness than Europeans, and educated and higher-income users require Trust to engage, whereas older adults and lower socioeconomic groups bypass it entirely, relying on perceived practical benefits (e.g., 24/7 availability, non-judgmental support). Our findings extend technology acceptance theory and inform the equitable design of emotional support AI.




\end{abstract}


\section{Introduction}

The integration of Large Language Models (LLMs) into our daily life has expanded beyond productivity tasks into the sensitive domain of emotional support \cite{anthropic2025affective, chatterji2025how}. Conversational agents are used increasingly by individuals seeking non-judgmental, always-available spaces to process their feelings, manage stress, and navigate personal relationships \cite{amat2026chatbots, chin2023potential, li2025human, chaudhry2024user}. 
Users have been treating LLMs as therapeutic companions \cite{chaudhry2024user, li2025human, hatch2025correction} because current models use highly fluent, anthropomorphic communication styles \cite{cheng2022human, epley2007seeing}. 
This reflects an important shift in human-AI interaction, moving from task-oriented usage toward affective digital companionship \cite{skjuve2021my, anthropic2025affective}. Understanding who trusts AI as a confidant and what drives trust is therefore a pressing scientific and societal concern, given that trust has emerged as a robust antecedent of AI adoption across domains \cite{yang2024chatbot, wu2025trust}. 

Yet research in this area lacks reliable tools to address this concern. A systematic review of
assessment methods for conversational agents used in mental wellbeing found that no validity
evidence was cited for more than half of the survey instruments used across the
literature \cite{jabir2023evaluating}, reflecting a broader absence of validated
psychometric instruments specifically designed to capture user perceptions of LLMs
in emotional support contexts. Traditional technology acceptance metrics compound
this problem by failing to capture what matters in emotional human-AI dialogue: the
social-affective dimensions (empathy, warmth) \cite{liu2018sympathy, rohden2026emotional} and the inclusion dimensions
(algorithmic bias, fairness) \cite{shin2020user, timmons2023call} that shape whether users feel safe enough to disclose
personal emotions \cite{marconi2026assessing, li2025human}. 


A second gap is the lack of 
large-scale sociodemographically diverse evidence demonstrating how different populations perceive, trust, and ultimately adopt AI for emotional support \cite{wang2024perceptions}. Technology acceptance and trust formation are not uniform; they are strongly influenced by cultural background, age, gender, education, and socioeconomic status \cite{kelly2023factors, venkatesh2003user, mendez2023you, bassignana2025ai, pew2023uncomfortable}.
When LLMs are deployed at scale (especially in sensitive contexts such as mental wellbeing)
, applying a ``one-size-fits-all'' assumption risks accentuating digital divides and delivering inadequate support to vulnerable groups \cite{bassignana2025ai, timmons2023call, omar2025sociodemographic}. 
Most existing studies on user perception of LLMs rely on small, homogeneous samples and rarely connect demographic characteristics to actual emotional support interactions \cite{wang2024perceptions}.
Understanding these differences is necessary to ensure that emotional support AI is designed equitably for a global user base \cite{timmons2023call}. 


To address these gaps, this paper pursues three research objectives:
\begin{enumerate}
    \item \textbf{Instrument development and validation}: Develop and validate a multidimensional psychometric scale grounded in a theoretically motivated conceptual model to assess user perceptions of LLMs for emotional support across social-affective, functional, and bias-related dimensions.
    \item \textbf{Structural modelling of trust and adoption}: Fit an SEM mapping the relationships between system attributes, Trust, Perceived Benefits, and Actual System Use.
    \item \textbf{Sociodemographic variation}: Conduct a Multi-Group Analysis (MGA) across gender, age, socioeconomic status, education level, and cross-national region to identify how trust formation and adoption patterns differ across user segments.
\end{enumerate}



Drawing on an international sample of 1,343 users across seven countries (USA, UK, Spain, Italy, France, Germany, Netherlands), this study makes three contributions. First, we introduce a validated
seven-construct psychometric instrument for assessing human--AI interaction in emotional support contexts, including a novel Perceived Bias scale. Second, the SEM reveals that Privacy, Personalization, and Humanlikeness (an emergent construct integrating affective empathy and anthropomorphism) drive Trust, while Perceived Bias degrades it. Third, the MGA reveals that trust formation and adoption logic vary systematically across all five demographic dimensions examined (gender, age, socioeconomic status, education, and cross-national region), demonstrating that trust in emotional support AI is not universal but depends critically on who the user is.


The rest of this paper is structured as follows. Section 2 reviews the relevant literature and develops the study hypotheses. Section 3 describes the instrument development, sampling strategy, and data collection procedures. Section 4 presents the data analysis and results. Section 5 discusses the theoretical and practical implications of the findings. Section 6 concludes the paper and outlines directions for future research.
\section{Literature Review and Hypotheses Development}

\subsection{LLMs for Emotional Support}

The emergence of large language models (LLMs) has shifted the usage of conversational agents for emotional support from narrow, rule-based scripts toward open-ended conversations \cite{anthropic2025affective}. LLMs are available around the clock, perceived as non-judgmental, confidential, and useful for self-reflection \cite{li2025human, chaudhry2024user}. They can generate therapeutic responses that are practically indistinguishable from those written by professional therapists and are sometimes rated as better aligned with core psychotherapy principles \cite{hatch2025correction}. At the same time, these systems exert behavioral influence: in a large randomized controlled trial, up to 79\% of participants followed advice received from an LLM after a single 20-minute session, even for high-stakes, hard-to-reverse decisions \cite{luettgau2025people}. Social-media discourse on TikTok is largely positive about LLMs as an ``emotional outlet'' and even as a substitute therapist \cite{haensch2025listens}. 

However, using LLMs for emotional support can be risky.
For example, a recent study showed that following AI advice produced no measurable well-being benefit two to three weeks later, suggesting that uptake does not translate to value \cite{luettgau2025people}. Worse, current models may show stigma toward conditions such as schizophrenia and alcohol dependence and respond inappropriately to critical situations (e.g., reinforcing delusional thinking through sycophancy) \cite{moore2025expressing}. These failures persist in larger and newer models, undermining therapeutic principles. Experts argue that ChatGPT may be therapeutic but lacks scientific validation as a psychotherapeutic tool, and it should complement human care rather than replace it \cite{pandya2024chatgpt}. Further concerns include shallow contextual understanding, weak session continuity, and lack of personalization \cite{chaudhry2024user}. There is also a reliance risk that discourages professional help-seeking \cite{li2025human}, and inadequate handling of abusive or boundary-testing input \cite{curry2019crowd}.

\subsection{Proposed Conceptual Model and Structural Configuration}

To map how users navigate these benefits and risks, we propose a multi-layered 
structural framework. This framework is built on the premise that system attributes do not trigger 
adoption directly; instead, their influence is mediated through two psychologically 
distinct evaluations that users form over time. These configure the model into three 
sequential layers:

\begin{enumerate}
    \item \textbf{System Attributes} (Independent Variables): The user's evaluations 
    of the system's functional and interactive characteristics, including Cognitive 
    Empathy, Affective Empathy, Anthropomorphism, Personalization, Privacy, and Perceived Bias.
    \item \textbf{Psychological Mediators}: A \textit{utilitarian pathway}, 
    operationalized as Perceived Benefits, and a \textit{relational pathway}, 
    operationalized as Trust, through which system attributes exert their influence 
    on behavior.
    \item \textbf{Behavioral Outcome} (Dependent Variable): Actual System Use, 
    operationalized as the self-reported frequency of LLM use for emotional support.
\end{enumerate}

\begin{figure}[t]
\centering
\includegraphics[width=0.7\textwidth]{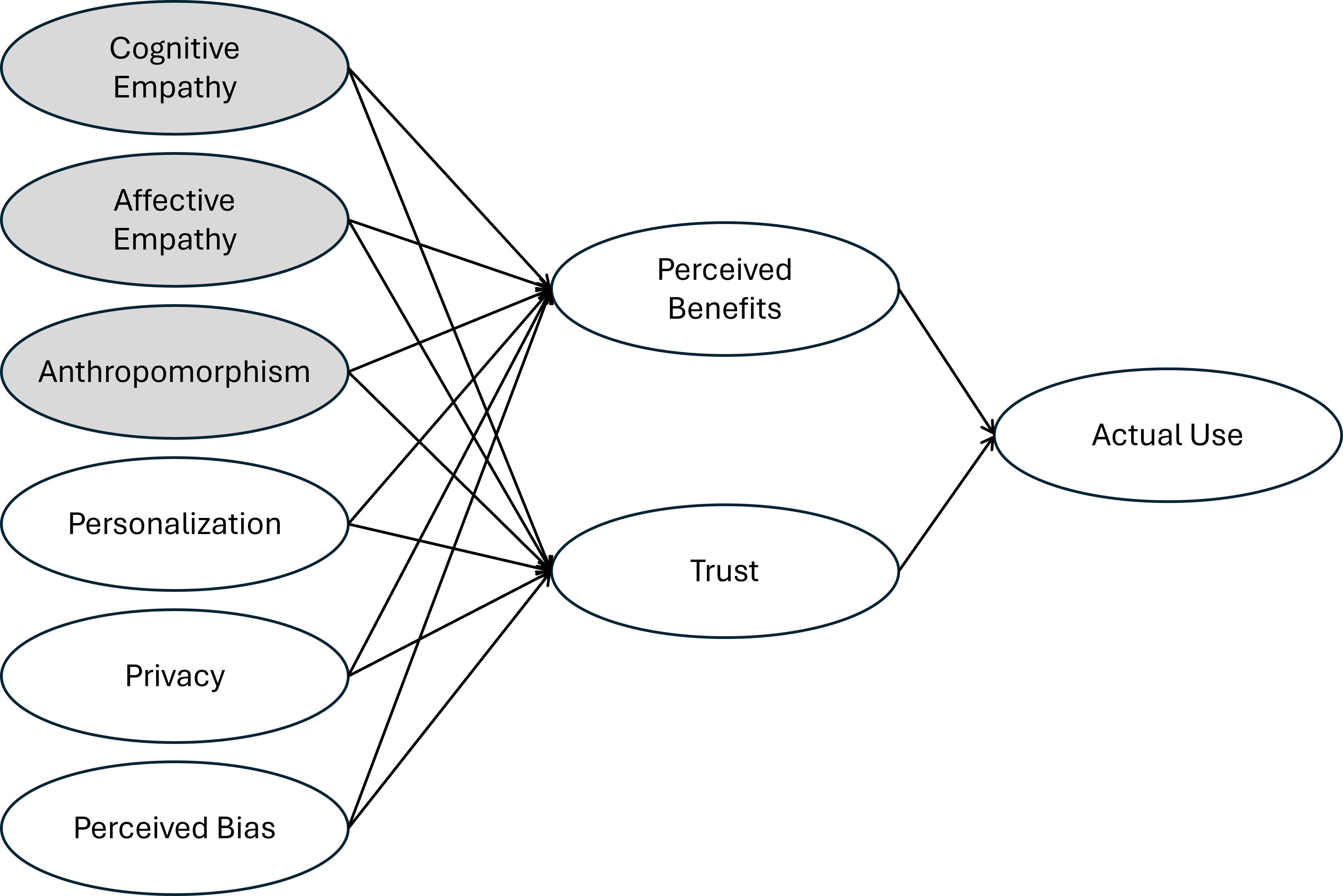} 
\caption{The Initial Conceptual Model outlining the originally hypothesized structural paths. Shaded components (Cognitive Empathy, Affective Empathy, and Anthropomorphism) represent different conceptual dimensions that were empirically merged into a single ``Humanlikeness'' construct following Exploratory Factor Analysis (EFA).}
\label{fig:initial_research_model}
\end{figure}

Figure~\ref{fig:initial_research_model} presents the full conceptual model. 
Section~\ref{sec:construct_definitions} defines each construct, grounds each 
pathway in the relevant theoretical literature, and develops the hypothesized 
structural relationships.

\subsection{Construct Definition and Hypotheses Development}
\label{sec:construct_definitions}

    \subsubsection{Actual System Use}
    
    In the Technology Acceptance Model (TAM), perceived usefulness (the degree to which a person believes a system enhances their performance) is the primary belief driving acceptance, with actual system use as the final behavioral outcome \cite{davis1989}. The Unified Theory of Acceptance and Use of Technology (UTAUT) identifies performance expectancy as the strongest predictor of behavioral intention and, subsequently, of use behavior \cite{venkatesh2003user}. We therefore adopt \textit{Actual System Use} as our outcome variable, measured as the self-reported frequency of LLM use for emotional support. 
    
     \subsubsection{Perceived Benefits}
    Users report that LLMs help them manage emotional difficulties, reflect on personal concerns, and express feelings without fear of judgment, alongside practical benefits such as 24/7 availability and lower cost compared to therapy \cite{li2025human, chaudhry2024user}. Social media discourse mirrors this benefit-driven appraisal, framing LLMs as an always-available ``emotional outlet'' \cite{haensch2025listens}, and benefit perceptions have been shown to carry acceptance in AI health services \cite{liu2022roles}. Inspired by TAM \cite{davis1989}, we extend perceived usefulness as \textit{Perceived Benefits} for the emotional support context, where the relevant ``performance'' is well-being. We hypothesize that the more benefits users perceive, the more likely they are to integrate the system into their support-seeking routines:

\begin{itemize}
     \item \textbf{H1:} Perceived Benefits positively influence the Actual System Use of LLMs for emotional support.
\end{itemize}

    \subsubsection{Trust}

   Trust is the willingness to be vulnerable to another party based on positive 
expectations of its conduct \cite{mayer1995integrative}, and it governs reliance on 
automation when complexity and stakes exceed what users can fully verify 
\cite{lee2004trust}. The trust literature distinguishes between initial trust, formed 
before or early in an interaction on the basis of perceived system characteristics, 
and experience-based trust, which develops and is refined through repeated use 
\cite{harrison2001trust}. The present study measures trust at a single point 
in time, capturing a global evaluation that reflects users' overall impression 
of the system. This cross-sectional 
measurement is consistent with how trust is captured in the majority of 
technology acceptance studies \cite{liu2022roles}, and it is appropriate for our 
research objective is to identify which system attributes predict trust levels among 
active users.

Whereas Perceived Benefits captures a utilitarian appraisal of what the system 
does, Trust reflects a qualitatively distinct evaluation based on interpersonal 
trust \cite{mayer1995integrative, harrison2001trust}: it concerns the 
perceived safety and integrity of the relationship rather than its functional output 
\cite{volpato2025trusting, komiak2006effects}. This distinction is particularly 
important in emotional support contexts, where users disclose intimate feelings and 
are exposed to harm beyond mere task failure. In such contexts, relational safety is 
a prerequisite for engagement that perceived usefulness alone cannot capture. Trust 
has emerged as a robust antecedent of AI adoption across domains, including reliance 
on conversational AI advice \cite{yazan2026personalized}, engagement with emotional 
support chatbots \cite{yang2024chatbot}, and intentions to adopt conversational 
agents for digital counseling \cite{wu2025trust}. We therefore model Trust as a 
second, complementary pathway to adoption alongside Perceived Benefits:

 \begin{itemize}
     \item \textbf{H2:} Trust positively influences the Actual System Use of LLMs for emotional support.
\end{itemize}

    \subsubsection{Empathy \& Anthropomorphism}
    
    The empathetic, non-judgmental tone is among the most valued qualities of LLM-based 
emotional support \cite{li2025human}. Empathetic LLM responses are rated as 
indistinguishable and sometimes better aligned with psychotherapy principles than 
therapist-written ones \cite{hatch2025correction}. Sympathetic and empathetic 
expressions from a health chatbot enhance perceptions of support 
\cite{liu2018sympathy}, and empathetic agents are perceived as warmer and produce 
higher satisfaction \cite{rohden2026emotional}. Empathy has two dimensions: 
a) cognitive empathy is the ability to comprehend and take the perspective of 
another's emotional state, and b) affective empathy is the sharing of and resonance 
with that state \cite{reniers2011qcae}.

Empathy is closely entangled with anthropomorphism, the attribution of humanlike 
mental states, intentions, and feelings to non-human agents \cite{epley2007seeing}. 
We treat cognitive empathy, affective empathy, and anthropomorphism as conceptually 
related but distinct attributes of the system, and enter them as separate constructs 
in the initial measurement model. On the perceived benefit side, empathy is the 
vehicle through which emotional support is delivered, and perceived empathy drives 
adoption of voice assistants while mitigating users' risk concerns 
\cite{coker2024alexa}. Warm, human-like conversation styles similarly lower 
perceived risk and strengthen perceived ties with chatbots \cite{xiao2025rethinking}. 
On the trust side, evidence links anthropomorphism to usefulness, trust, and usage 
intention of service AI \cite{blut2021understanding}, and anthropomorphism predicts 
trust in smart healthcare services \cite{liu2022roles}.

However, the three constructs share substantial conceptual overlap: all three concern 
the degree to which users perceive the system as human-like in its understanding and 
expression of emotion. This raises the empirical question of whether users actually 
experience them as separable dimensions or as a unified quality. We leave this as an 
open question to be addressed by the Exploratory Factor Analysis (Section 
\ref{sec:efa}), which will determine whether the data support their treatment as 
distinct constructs or warrant their consolidation. Pending that test, we 
provisionally hypothesize their joint influence:

 \begin{itemize}
     \item \textbf{H3a:} Cognitive empathy, affective empathy, and anthropomorphism positively influence Perceived Benefits of LLMs for emotional support.
     \item \textbf{H3b:} Cognitive empathy, affective empathy, and anthropomorphism positively influence Trust in LLMs for emotional support.
\end{itemize}
    \subsubsection{Personalization}
    
    Personalization refers to tailoring system responses to an individual's needs, preferences, and interaction history \cite{hawkins2008tailoring}. A systematic review found that personalization of healthcare conversational agents improves satisfaction and engagement \cite{kocaballi2019personalization}, and personalized AI-generated health advice increases cognitive and emotional trust, which in turn drives adoption intention \cite{qin2024examining}. Perceived personalization likewise increases cognitive and emotional trust in recommendation agents \cite{komiak2006effects}, and personalization predicts both trust and behavioral intention in public acceptance of smart healthcare services \cite{liu2022roles}. Users cite generic, one-size-fits-all responses and the lack of memory across sessions as principal shortcomings of mental health chatbots \cite{chaudhry2024user}, investing substantial effort in personas and iterative prompting to coax ChatGPT into responses tailored to their emotional needs \cite{li2025human}. A system that remembers, adapts, and responds to the individual should accordingly be perceived as both more beneficial and more trustworthy:

      \begin{itemize}
     \item \textbf{H4a:} Personalization positively influences Perceived Benefits of LLMs for emotional support.
     \item \textbf{H4b:} Personalization positively influences Trust in LLMs for emotional support.
\end{itemize}

    \subsubsection{Privacy}
    
    Privacy concerns are the most negatively discussed topic in social media discourse on LLMs as mental health tools \cite{haensch2025listens}, and perceived privacy risk is a central determinant of how users evaluate chatbots \cite{xiao2025rethinking}. Feeling safe to disclose is a precondition for the benefits of emotional support to materialize, since open emotional expression without fear of judgment or exposure is itself a core perceived benefit \cite{li2025human, chaudhry2024user}. For trust, the link is well established: perceived loss of privacy significantly erodes trust in smart healthcare services \cite{liu2022roles}, identifying privacy as a key system-level determinant of trust in AI. We expect users who perceive their interactions as private, safe, and transparently handled to perceive the system as more beneficial and place more trust in it:

      \begin{itemize}
     \item \textbf{H5a:} Privacy positively influences Perceived Benefits of LLMs for emotional support.
     \item \textbf{H5b:} Privacy positively influences Trust in LLMs for emotional support.
\end{itemize}

    \subsubsection{Perceived Bias}

    \label{sec:perceivedbias}
    
    AI systems in mental health amplify societal biases, and researchers have called for a systematic assessment of such bias toward cultural, gender, and identity groups as a prerequisite for equitable deployment \cite{timmons2023call}. Models systematically over-refer LGBTQIA+ and minority patients in medical decision-making \cite{omar2025sociodemographic}, produce gender-stereotyped emotional responses \cite{grogan2025ai}, display empathy gaps toward social out-groups \cite{hou2025language}, and express stigma toward conditions such as schizophrenia and alcohol dependence \cite{moore2025expressing}. When users perceive algorithmic decisions as unfair, they judge the system as less useful and less trustworthy \cite{shin2020user}, and biased or stigmatizing responses undermine emotional support \cite{moore2025expressing}. However, how users \textit{perceive} bias remains understudied: Wang et al. \cite{wang2024perceptions} reviewed 15 bias studies and found that only two provided explicit definitions of bias-related terms, and none used a validated instrument for capturing users' subjective experience of bias. They note that most studies employ performance metrics such as accuracy and coherence to capture bias. However, measuring bias perceptions requires dedicated scales and demographically diverse samples to be properly conceptualized and measured \cite{wang2024perceptions}. We therefore introduce the \textit{Perceived Bias} construct. It captures perceptions regarding whether the system makes prejudiced assumptions about one's cultural or religious background, gender, or sexual orientation, pushes a single viewpoint on well-being, and fails to acknowledge its own limitations and cultural gaps. Unlike the preceding system attributes, we hypothesize negative relationships:

\begin{itemize}
     \item \textbf{H6a:} Perceived Bias negatively influences Perceived Benefits of LLMs for emotional support.
     \item \textbf{H6b:} Perceived Bias negatively influences Trust in LLMs for emotional support.
\end{itemize}
     
Based on the literature review and the hypotheses, we present the relationships between these distinct behavioral dimensions in the Initial Conceptual Model (Figure~\ref{fig:initial_research_model}).

\subsection{Sociodemographic Perspectives in AI Interaction}
\label{sec:sociodemographic_background}

While the hypotheses above outline our baseline structural model, acceptance of AI in emotional support is not uniform across the population. Younger adults, men, and those with higher levels of education are more receptive to using AI, and prior familiarity matters greatly, with discomfort dropping from 70\% among those who had never heard of AI to 50\% among those who had heard a lot about it \cite{pew2023uncomfortable}. Individual differences in personality and technology readiness likewise shape how LLM-generated advice is received, with agreeableness and technological insecurity predicting users preferring AI for sensitive, private topics such as mental health \cite{wester2024exploring}. Age also shapes how LLM responses are perceived: Generation Z rated avoidance strategies of LLMs for verbal abuse lower than other groups, while older participants viewed humorous responses to harassment as highly inappropriate \cite{curry2019crowd}. Beyond user perceptions, models themselves treat sociodemographic groups unequally. In medical decision-making, LLMs systematically over-refer LGBTQIA+ patients for mental-health assessment and inflate triage urgency for Black, unhoused, and transgender patients \cite{omar2025sociodemographic}. Assigning gendered relationship personas elicits stereotyped responses: models associate anger with male personas and distress with female ones \cite{plaza-del-arco-etal-2024-angry,grogan2025ai}. 

Existing work tends to fall into one of two camps: studies of model bias that probe outputs with synthetic identities but do not engage real users \cite{omar2025sociodemographic, hou2025language, grogan2025ai}, or studies of user perception and acceptance that rarely connect demographics to actual emotional-support interactions. Reviews of the perception literature explicitly note that most studies use small samples and fail to collect demographic information \cite{wang2024perceptions}. Consequently, little is known about how sociodemographic characteristics (e.g., gender, age, education, and socioeconomic status) jointly shape acceptance, emotional disclosure, and privacy expectations when people seek emotional support from LLMs. Our study addresses this gap by conducting a multi-group analysis of emotional-support interactions with LLMs across sociodemographic strata. 

In addition to the proposed sociodemographic differences, culture (based on the Inglehart-Welzel cultural map \cite{inglehart2005modernization}, and Hofstede's dimensions \cite{hofstede2001cultures}) shapes perceptions of non-human entities. These frameworks originate in cross-cultural psychology, sociology, and political science and have predicted national differences in technology acceptance, privacy concern, and trust in automation \cite{bellman2004international, im2011international}. For example, the Anglosphere has some of the world's highest individualism scores and low uncertainty avoidance, a profile associated with early adoption of novel technologies and a readiness to extend social responses to non-human agents \cite{srite2006role}. Mediterranean countries score high on uncertainty avoidance, suggesting increased sensitivity to cues that a counterpart is prejudiced or unfair \cite{putnam1993making, banfield1958moral, hofstede2001cultures}. Western Europe is characterized as consensus-oriented \cite{hofstede2010cultures}, implying that trust might follow demonstrated quality rather than affective, humanlike cues. 

The three regional clusters used in our multi-group analysis (Anglosphere: UK, USA; 
Southern Europe: Italy, Spain; Western/Central Europe: France, Germany, Netherlands) 
reflect a combination of cultural and geographic proximity. We acknowledge that this 
grouping is an approximation: the Netherlands, for instance, scores high on 
individualism similarly to the Anglosphere \cite{hofstede2001cultures}, and could 
plausibly be grouped differently on purely cultural grounds. However, it shares 
geographic, institutional, and regulatory context with its Western European neighbors 
(including GDPR jurisdiction and broadly consensus-oriented political culture). The clustering 
should therefore be interpreted as a pragmatic grouping that balances cultural 
theory with geographic coherence and statistical feasibility, and findings for the 
Western/Central Europe cluster should be interpreted with this caveat in mind.

Given the early stage of the literature on sociodemographic variations in adoption and trust formation towards AI for emotional support, we adopt an exploratory approach to the multi-group analysis. Rather than formulating directional hypotheses, we aim to answer the following research question: 
\begin{itemize}
    \item How do the structural relationships between system attributes, Trust, Perceived Benefits, and Actual Usage vary across distinct sociodemographic groups?
\end{itemize}

\section{Methodology}

We adopted a cross-sectional, quantitative survey design, developing and validating a psychometric instrument for assessing user perceptions of LLMs in affective contexts, collecting data from an international sample of active users across seven countries, and analyzing the results through a multi-stage approach:
Exploratory Factor Analysis (EFA) to discover latent structure, Confirmatory Factor Analysis (CFA) to validate it, Structural Equation Modeling (SEM) to test the 
hypothesized relationships between constructs, and Multi-Group Analysis (MGA) to 
examine how those relationships vary across sociodemographic segments.

\subsection{Instrument Development}

We developed the survey instrument through a systematic process involving initial 
item pool generation, expert panel review, pilot testing, and translation workflows. 
We generated the initial item pool based on an extensive review of the Human Computer Interaction (HCI) literature to identify relevant behavioral and evaluative dimensions. Items 
were mapped across nine distinct latent variables (defined in Section \ref{sec:construct_definitions}): Trust, Privacy, 
Personalization, Anthropomorphism, Cognitive Empathy, Affective Empathy, Perceived 
Bias, Perceived Benefits, and Actual System Use. We adapted the items either from 
established psychometric frameworks --- such as the Technology Acceptance Model 
(TAM) \cite{davis1989}, the Questionnaire of Cognitive and Affective Empathy (QCAE) 
\cite{reniers2011qcae}, and recent HCI scales \cite{marimon2024, schmidmaier2024} 
--- or formulated by the authors to fit the specific context of LLM use for emotional 
and mental wellbeing support. 

Regarding Perceived Bias, while psychometric scales for related constructs such as algorithmic fairness and 
AI bias exists; none were designed to capture the specific dimensions of bias 
perception most relevant to emotional support contexts — including assumptions about 
cultural background, gender, sexual orientation, and viewpoint plurality. We 
therefore developed the Perceived Bias items ourselves, drawing on the literature 
reviewed in Section~\ref{sec:perceivedbias}, to ensure coverage of several bias 
dimensions relevant to users disclosing personal information to an LLM.

To ensure content validity, the initial English item pool was evaluated by a 
cross-disciplinary panel of 20 experts from our target countries, including 
specialists in survey methodology, Natural Language Processing (NLP), and HCI. 
Experts reviewed items for clarity, relevance, and contextual appropriateness, and 
we used their feedback to refine question wording and improve readability. For 
instance, we softened absolute or deterministic statements into subjective, 
perception-focused phrasing (e.g.,~modifying ``I can express my feelings'' to ``I 
feel that I can express my feelings''). Subsequently, we conducted a pilot study to 
assess item comprehension and estimate completion time. We used the recorded average 
completion time of 8--9 minutes to configure platform parameters and 
calculate participant compensation.

Because the target audience spanned multiple regions, we executed a formal translation process. The original English survey was automatically translated into Spanish, 
French, Italian, German, and Dutch via Qualtrics and subsequently reviewed and 
edited by native-speaking experts from our panel to ensure conceptual equivalence 
across language versions.

We present the final survey instrument in Appendix \ref{tab:survey_items}. 

\subsection{Participants}

We implemented an international sampling strategy through a Qualtrics panel in 
November 2025 to examine demographic and cultural variation in user perceptions of 
LLMs for emotional support. To ensure our sample reflected populations actively 
engaging with these systems, we recruited participants from seven countries 
representing the highest global shares of ChatGPT visitors in North America and 
Europe \cite{firstpagesage2026chatgpt}: the United States (17.1\%), France (4.3\%), 
Spain (3.7\%), the United Kingdom (2.7\%), Italy (2.5\%), Germany (2.4\%), and the 
Netherlands (1.1\%). Quota sampling was enforced by the Qualtrics platform to target 
a balanced baseline across countries ($N = 200$ per country) and an even gender 
balance within each national cohort.

\subsection{Data collection}

\begin{table}[htbp]
    \centering
    \small
    \setlength{\tabcolsep}{5pt} 
    \caption{Users of LLMs for emotional support. Breakdown of sociodemographic characteristics across countries (counts and overall percentage). The most common group in each category is boldfaced per column. SES refers to the Socioeconomic status. UK = United Kingdom, USA = United States of America, NL = Netherlands, ESP = Spain, FRA = France, GER = Germany, ITA = Italy.}
    \label{tab:Demographics}
    \begin{tabular}{@{}lccccccccc@{}}
    \toprule
    \textbf{} & \textbf{UK} & \textbf{USA} & \textbf{NL} & \textbf{ESP} & \textbf{FRA} & \textbf{GER} & \textbf{ITA} & \textbf{Total} & \textbf{\%} \\
    \midrule
    \textbf{Gender} &  &  &  &  &  &  &  &  & \\
    \quad Men & 90 & \textbf{92} & 95 & 99 & 91 & 94 & \textbf{99} & 660 & 49.14\% \\
    \quad Women & \textbf{94} & 92 & \textbf{100} & \textbf{100} & \textbf{98} & \textbf{98} & 99 & \textbf{681} & \textbf{50.71\%} \\
    \quad Non-binary/Other & 0 & 0 & 0 & 1 & 1 & 0 & 0 & 2 & 0.15\% \\
    \addlinespace
    \textbf{Age} &  &  &  &  &  &  &  &  & \\
    \quad Under 18 & 0 & 0 & 0 & 0 & 1 & 1 & 0 & 2 & 0.15\% \\
    \quad 18--24 & 8 & 4 & 16 & 22 & 21 & 22 & 16 & 109 & 8.12\% \\
    \quad 25--34 & \textbf{117} & 41 & 44 & 53 & \textbf{60} & 39 & \textbf{55} & \textbf{409} & \textbf{30.45\%} \\
    \quad 35--44 & 44 & \textbf{87} & \textbf{58} & 42 & 45 & \textbf{51} & 43 & 370 & 27.55\% \\
    \quad 45--54 & 10 & 27 & 38 & \textbf{65} & 33 & 28 & 48 & 249 & 18.54\% \\
    \quad 55--64 & 4 & 13 & 22 & 16 & 17 & 34 & 27 & 133 & 9.90\% \\
    \quad 65+ & 1 & 12 & 17 & 2 & 13 & 17 & 9 & 71 & 5.29\% \\
    \addlinespace
    \textbf{Education} &  &  &  &  &  &  &  &  & \\
    \quad Primary school & 0 & 1 & 2 & 2 & 5 & 4 & 3 & 17 & 1.27\% \\
    \quad HS & 9 & 26 & 36 & 18 & \textbf{61} & \textbf{59} & \textbf{68} & 277 & 20.63\% \\
    \quad Some college & 9 & 13 & 2 & 8 & 11 & 15 & 25 & 83 & 6.18\% \\
    \quad Associate degree & 0 & 10 & \textbf{48} & 31 & 39 & 1 & 9 & 138 & 10.28\% \\
    \quad BSc & 46 & 52 & 48 & 59 & 35 & 30 & 38 & \textbf{308} & \textbf{22.93\%} \\
    \quad MSc & \textbf{58} & \textbf{63} & 41 & 12 & 30 & 27 & 32 & 263 & 19.58\% \\
    \quad PhD & 45 & 10 & 5 & 8 & 2 & 4 & 4 & 78 & 5.81\% \\
    \quad Professional degree & 16 & 9 & 13 & \textbf{62} & 6 & 50 & 19 & 175 & 13.03\% \\
    \addlinespace
    \textbf{SES} &  &  &  &  &  &  &  &  & \\
    \quad 1 (Worst off) & 1 & 2 & 2 & 1 & 1 & 3 & 1 & 11 & 0.82\% \\
    \quad 2 & 1 & 1 & 3 & 1 & 6 & 2 & 3 & 17 & 1.27\% \\
    \quad 3 & 1 & 7 & 3 & 5 & 16 & 12 & 4 & 48 & 3.57\% \\
    \quad 4 & 5 & 11 & 11 & 22 & 31 & 12 & 22 & 114 & 8.49\% \\
    \quad 5 & 12 & 21 & 16 & 30 & 35 & 28 & 39 & 181 & 13.48\% \\
    \quad 6 & 13 & 22 & 23 & \textbf{53} & \textbf{46} & \textbf{39} & \textbf{44} & 240 & 17.87\% \\
    \quad 7 & \textbf{45} & 33 & \textbf{71} & 46 & 32 & 39 & 40 & \textbf{306} & \textbf{22.78\%} \\
    \quad 8 & 44 & \textbf{38} & 54 & 26 & 10 & 34 & 33 & 239 & 17.80\% \\
    \quad 9 & 29 & 27 & 8 & 12 & 5 & 14 & 6 & 101 & 7.52\% \\
    \quad 10 (Best off) & 33 & 22 & 4 & 4 & 8 & 9 & 6 & 86 & 6.40\% \\
    \midrule
     & $184$ & $184$ & $195$ & $200$ & $190$ & $192$ & $198$ & $1{,}343$ & $100.00\%$ \\
    \bottomrule
    \end{tabular}
\end{table}

We collected the data through an online questionnaire distributed via Qualtrics\footnote{\url{https://www.qualtrics.com/}}. 
Before starting the survey, we presented participants with an onboarding screen to 
obtain formal digital informed consent. This screen outlined the study objectives, 
guaranteed total anonymity under GDPR compliance, and established a shared 
conceptual baseline by providing examples of ``AI agents'' (e.g., ChatGPT, Gemini, 
Claude) and defining ``emotional support and mental wellbeing'' as using these tools 
for managing stress, processing emotions, or personal reflection.

Participants then completed a sociodemographic block capturing standard variables (gender, age, highest education level completed)
alongside a subjective assessment of Socioeconomic status (SES). More specifically, we used the MacArthur Scale 
of Subjective Social Status \cite{adler2000relationship}, where participants rank their perceived social standing by marking a rung on a 10-step ladder. Unlike traditional (objective) measures like income or education, it assesses how people view their own social rank relative to others, making it particularly adequate for 
cross-country analyses. A mandatory gating filter question then classified 
respondents as users (those who had used AI chatbots for emotional support) or 
non-users. The structural analyses reported in this study focus exclusively on the 
user cohort.

Prior to data collection, we obtained formal institutional ethics approval (See Ethics approval statement). To ensure data 
integrity, we applied strict quality filters to the initial pool of $N = 5{,}319$ raw 
responses. Qualtrics platform filters automatically excluded suspected bot activity, 
duplicate submissions, incomplete responses, and straight-lining. We additionally 
excluded participants with completion times faster than 30\% of the median to 
eliminate speeders. 
We randomly embedded two attention-check items within the 
psychometric matrices: the first required selection of a specific neutral value 
(\textit{``Please select `Neither Agree nor Disagree' for this statement to confirm 
you are paying attention''}), and the second required an extreme validation value 
(\textit{``Please select `Strongly Agree' to confirm you are paying attention''}). 
Failure to correctly answer either check resulted in exclusion from the dataset. Following all quality 
filters, $N = 4{,}641$ valid responses remained across users and non-users combined.
We measured each substantive item on a 5-point Likert scale ranging from 
\textit{Strongly disagree} to \textit{Strongly agree}. Because the survey interface 
required complete responses before final submission, the final dataset contained 
zero missing values. Of the $N = 4{,}641$ valid responses, $N = 1{,}343$ were from active users of 
AI for emotional support and constitute the analytical sample for this study. The 
proportion of users varied notably across countries, ranging from 20.2\% in France 
and 20.8\% in the Netherlands to 49.0\% in Spain and 59.0\% in the United Kingdom, 
with the United States (31.5\%), Germany (24.6\%), and Italy (29.2\%) falling 
between these extremes. The sociodemographic characteristics of the active users sample are 
summarized in Table~\ref{tab:Demographics}.

\subsection{Analytical Strategy}

Data analysis proceeded in four stages. First, we did an Exploratory Factor Analysis (EFA) \cite{watkins2021step} to identify the 
underlying latent structure of the item pool. We conducted two separate EFAs using 
Varimax rotation with Kaiser normalization: one isolating the independent variable 
items, and one processing the response and mediating construct items.

Second, we subjected the retained factor structure to CFA in AMOS \cite{collier2020applied}, evaluating model fit via 
standard indices including CFI, RMSEA, and SRMR. Third, we estimated a SEM \cite{ullman2012structural} testing the 
hypothesized paths from system attributes through Trust and Perceived Benefits to 
Actual System Use. We estimated indirect effects via bootstrapping with 2,000 
iterations and bias-corrected 95\% confidence intervals.

Fourth, we conducted a Multi-Group Analysis (MGA) across five demographic 
dimensions: gender, age group, cross-national region, socioeconomic status, and education level. Before 
comparing structural paths across groups, we established measurement invariance  
for each configuration via chi-square difference tests ($\Delta\chi^2$) comparing 
constrained and unconstrained models. We identified significant cross-group path differences using Critical Ratios for differences between parameters, with $|C.R.| > 
1.96$ ($p < .05$) as the threshold for significance.

\section{Data Analysis and Results}

We split the clean user dataset ($N = 1,343$) into two randomly generated halves via SPSS (Statistical Package for the Social Sciences) \cite{hinton2014spss}. We used the first subsample ($n = 671$) to conduct the Exploratory Factor Analyses (EFA) to discover and refine latent configurations. We used the second subsample ($n = 672$) for subsequent CFA. For the SEM, we used the entire sample.

\subsection{Exploratory Factor Analysis (EFA)}
\label{sec:efa}

We conducted two separate EFAs in SPSS using Varimax rotation with Kaiser normalization. The first EFA isolated the self-reported independent variable fields, and the second processed the response constructs. We applied a strict factor loading cutoff threshold of $.55$ across both runs. We systematically removed items showing cross-loadings above $.5$ or failing to clear the $.55$ threshold to ensure structural integrity.

Data suitability parameters for both factor models showed excellent sampling metrics. The independent factor run achieved a Kaiser-Meyer-Olkin (KMO) measure of sampling adequacy of $.973$, with an approx.\ Bartlett’s Test of Sphericity $\chi^2(496) = 24370.276$, $p < .001$. The response factor run achieved a KMO value of $.965$, with a Bartlett’s Test of Sphericity $\chi^2(153) = 13493.687$, $p < .001$. These metrics confirm that the indicators share sufficient common variance to proceed with component extraction.

\subsubsection{The ``Humanlikeness'' Construct}
A structural realignment emerged during the evaluation of the independent variable 
matrix that addresses the open question raised in 
Section~\ref{sec:construct_definitions} (whether users actually
experience Affective Empathy, Cognitive Empathy, and Anthropomorphism as separable dimensions or as a unified quality). Indicators originally designed to capture 
conceptually distinct dimensions (Anthropomorphism, Cognitive Empathy, and 
Affective Empathy) all loaded strongly onto a single unified factor, with loading 
values between $.586$ and $.742$ (see Table~\ref{tab:efa_independent}).

This result suggests that, in the context of LLMs used for emotional 
support, users do not experience these attributes as separable dimensions. Rather 
than distinguishing between a system's human-like mimicry (Anthropomorphism), its 
rational perspective-taking (Cognitive Empathy), and its emotional resonance 
(Affective Empathy), users appear to form a single holistic impression of whether 
the system feels human-like in its interaction. This is consistent with the 
conceptual overlap noted in Section~\ref{sec:construct_definitions} and with 
qualitative evidence that users attend primarily to interactional qualities such as 
conversational flow and authenticity rather than to theoretically separable facets 
of social cognition \cite{schimmelpfennig2025humanlike}. Accordingly, we consolidated the three 
sub-scales into a single merged factor, which we name 
\textit{Humanlikeness}, to represent this integrated user experience. Hypotheses 
H3a and H3b are retained under this consolidated construct.


\begin{figure}[t]
\centering
\includegraphics[width=0.7\textwidth]{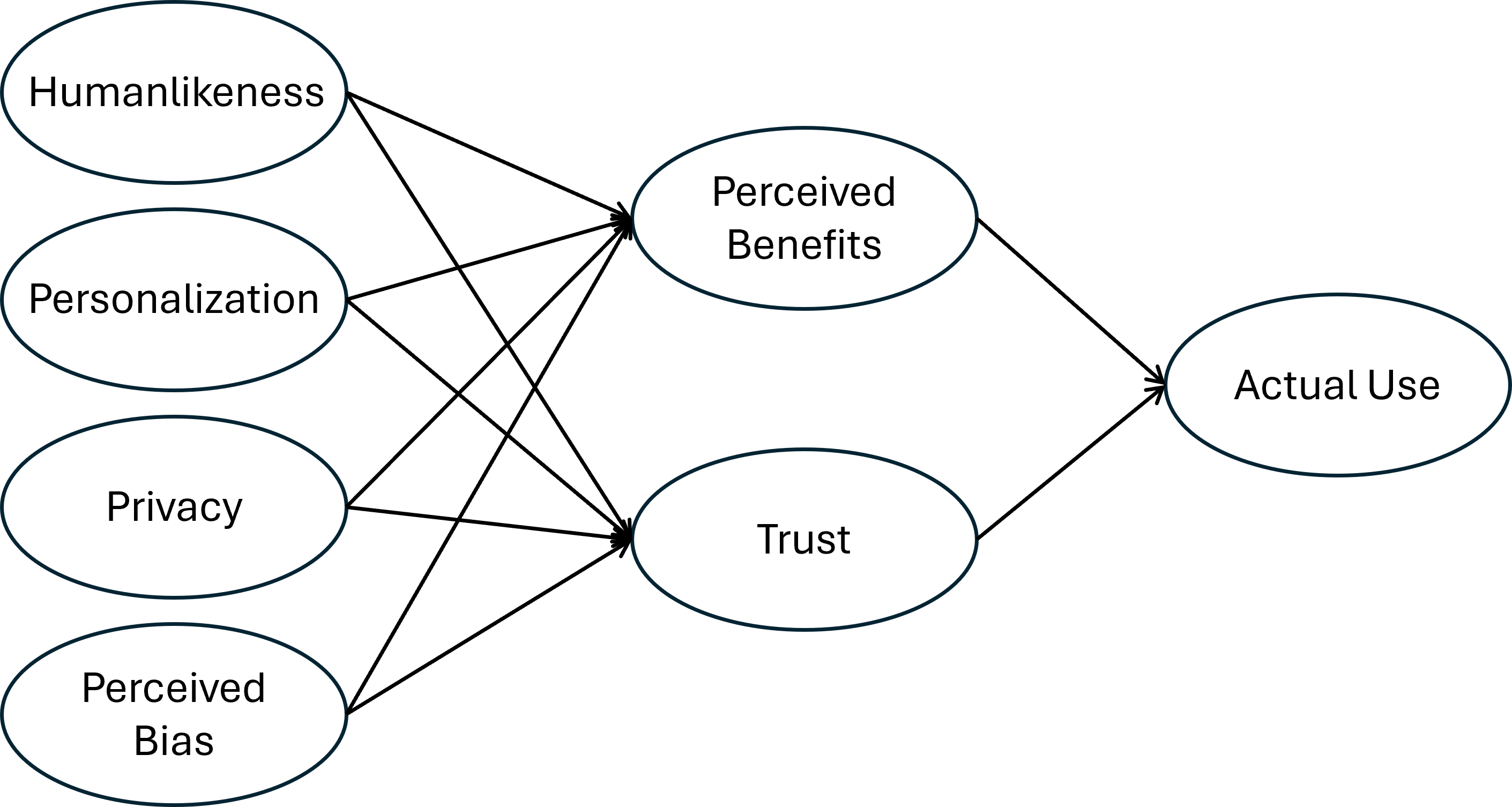}
\caption{The final Research Model mapping the effects of system attributes (Humanlikeness, Personalization, Privacy, and Perceived Bias) on the mediating variables (Perceived Benefits and Trust) and the final outcome variable (Actual Use) for LLM emotional support.}
\label{fig:research_model}
\end{figure}

\subsubsection{Item Elimination and Factor Structure Purification}
To ensure unidimensionality and prevent overlapping variance before running the structural paths, we dropped several items because they failed to meet the $.55$ loading criteria or exhibited cross-loading issues. 

In the independent variables model, we removed four items: \texttt{COG2} (.519), \texttt{ANT1} (.453), \texttt{BIA6} (.532), and \texttt{BIA5} (.498). Additionally, we removed \texttt{PRI6} (.454), \texttt{PER1} (.514), \texttt{PRI1} (.395), and \texttt{ANT6} (.057) due to low or split-factor behavior. 

In the response variables matrix, we dropped three items from the finalized evaluation pool because they failed to clear the $.55$ threshold: \texttt{BEN2\_3} (.518), \texttt{BEN1\_1} (.461), and \texttt{BEN1\_2} (.329). Additionally, we excluded \texttt{BEN2\_6} (.474) because it cross-loaded significantly onto actual system use (.521). 

Following these EFA refinements, we systematically removed two additional items during the subsequent multi-group CFA validation phase to maximize multi-group discriminant validity: \texttt{BIA7} and \texttt{BEN2\_4}. The full rotated factor structures, loading components, and item retention results are summarized in Tables \ref{tab:efa_independent} and \ref{tab:efa_response} in Appendix \ref{app:EFA_tables}.


\subsection{Confirmatory Factor Analysis (CFA)}

Following the dimension refinement completed during the exploratory phase, we conducted a CFA using AMOS on the second random split-sample ($n = 672$) to evaluate the psychometric properties of the measurement model. We specified the structural framework with seven distinct latent constructs based on the streamlined EFA indicators: Humanlikeness (HUM), Privacy (PRI), Personalization (PER), Perceived Bias (BIA), Trust (TRU), Perceived Benefits (BEN), and Actual System Use (USE).

\subsubsection{Measurement Model Fit}
The evaluation of the baseline measurement model revealed an excellent statistical representation of the data. Model fit indices safely cleared their respective standard psychometric thresholds. The model fit indicators are summarized in Table \ref{tab:cfa_model_fit}.

The normative chi-square was within acceptable parameters ($\chi^2/df = 4.117$, with $\chi^2 = 2359.221$ and $df = 573$). Reflecting incremental fit parameters, the Comparative Fit Index (CFI = .940) and the Tucker-Lewis Index (TLI = .934) both demonstrated a robust model configuration exceeding the standard $.90$ benchmark. Regarding error-based measures, the Root Mean Square Error of Approximation (RMSEA) was $.048$ ($90\%$ CI: $[.046, .050]$), and the Standardized Root Mean Square Residual (SRMR) was $.038$. Both metrics comfortably met strict publication criteria ($\le .05$), proving the theoretical configuration closely maps actual user response structures.

\begin{table*}[t]
\centering
\scriptsize
\caption{Measurement Model Goodness-of-Fit Indices}
\label{tab:cfa_model_fit}
\begin{tabular}{p{4.5cm}ccrc}
\toprule
\textbf{Fit Index Measure} & \textbf{Recommended Threshold} & \textbf{Value} & \textbf{Assessment} \\
\midrule
Normed Chi-Square ($\chi^2/df$) & $\le 5.0$ (Acceptable); $\le 3.0$ (Good) & 4.117 & Acceptable fit \\
Comparative Fit Index (CFI) & $\ge .90$ & .940 & Good fit \\
Tucker-Lewis Index (TLI) & $\ge .90$ & .934 & Good fit \\
Normed Fit Index (NFI) & $\ge .90$ & .922 & Good fit \\
Incremental Fit Index (IFI) & $\ge .90$ & .940 & Good fit \\
RMSEA & $\le .05$ (Excellent); $\le .08$ (Acceptable) & .048 & Excellent fit \\
SRMR & $\le .08$ & .038 & Excellent fit \\
\bottomrule
\end{tabular}
\end{table*}

\subsubsection{Convergent Validity and Reliability}
To establish convergent validity, we examined four evaluation criteria: standardized item factor loadings ($\lambda$), Cronbach's alpha ($\alpha$), Composite Reliability ($CR$), and Average Variance Extracted ($AVE$). The psychometric values extracted from the baseline sample matrices are detailed in Tables \ref{tab:validity_iv} and \ref{tab:validity_dv}.

\begin{table}[htbp]
\centering
\small
\caption{Measurement Model Validity and Reliability: Independent Variables}
\label{tab:validity_iv}
\begin{tabular}{@{}lcccc@{}}
\toprule
\textbf{Construct / Item} & \textbf{Loading ($\lambda$)} & \textbf{$\alpha$} & \textbf{$CR$} & \textbf{$AVE$} \\
\midrule
\textbf{Humanlikeness (HUM)} & & .926 & .926 & .534 \\
\quad ANT2 & .718 & & & \\
\quad ANT3 & .723 & & & \\
\quad ANT4 & .741 & & & \\
\quad ANT5 & .771 & & & \\
\quad AFF1 & .710 & & & \\
\quad AFF2 & .737 & & & \\
\quad AFF3 & .673 & & & \\
\quad AFF4 & .739 & & & \\
\quad COG1 & .783 & & & \\
\quad COG3 & .724 & & & \\
\quad COG4 & .711 & & & \\
\midrule
\textbf{Privacy (PRI)} & & .896 & .899 & .689 \\
\quad PRI2 & .818 & & & \\
\quad PRI3 & .839 & & & \\
\quad PRI4 & .876 & & & \\
\quad PRI5 & .785 & & & \\
\midrule
\textbf{Personalization (PER)} & & .766 & .772 & .530 \\
\quad PER2 & .752 & & & \\
\quad PER3 & .707 & & & \\
\quad PER4 & .724 & & & \\
\midrule
\textbf{Perceived Bias (BIA)} & & .842 & .841 & .515 \\
\quad BIA1 & .730 & & & \\
\quad BIA2 & .727 & & & \\
\quad BIA3 & .705 & & & \\
\quad BIA4 & .718 & & & \\
\quad BIA8 & .707 & & & \\
\bottomrule
\multicolumn{5}{@{}p{\linewidth}@{}}{\footnotesize \textit{Note:} $\alpha$ = Cronbach's Alpha, $CR$ = Composite Reliability, $AVE$ = Average Variance Extracted.}
\end{tabular}
\end{table}

\begin{table}[htbp]
\centering
\small
\caption{Measurement Model Validity and Reliability: Mediator and Response Variables}
\label{tab:validity_dv}
\begin{tabular}{@{}lcccc@{}}
\toprule
\textbf{Construct / Item} & \textbf{Loading ($\lambda$)} & \textbf{$\alpha$} & \textbf{$CR$} & \textbf{$AVE$} \\
\midrule
\textbf{Trust (TRU)} & & .859 & .860 & .605 \\
\quad TRU1 & .800 & & & \\
\quad TRU2 & .748 & & & \\
\quad TRU3 & .749 & & & \\
\quad TRU5 & .813 & & & \\
\midrule
\textbf{Perceived Benefits (BEN)} & & .871 & .859 & .505 \\
\quad BEN2\_1 & .656 & & & \\
\quad BEN1\_4 & .738 & & & \\
\quad BEN2\_2 & .673 & & & \\
\quad BEN1\_5 & .720 & & & \\
\quad BEN1\_3 & .698 & & & \\
\quad BEN2\_5 & .773 & & & \\
\midrule
\textbf{Actual System Use (USE)} & & .869 & .869 & .688 \\
\quad USE1 & .827 & & & \\
\quad USE2 & .841 & & & \\
\quad USE3 & .821 & & & \\
\bottomrule
\multicolumn{5}{@{}p{\linewidth}@{}}{\footnotesize \textit{Note:} $\alpha$ = Cronbach's Alpha, $CR$ = Composite Reliability, $AVE$ = Average Variance Extracted.}
\end{tabular}
\end{table}

Standardized factor loadings for all retained indicator variables were strong, ranging between $.656$ and $.876$, which ensures that individual items share significant common variance with their parent constructs. Internal consistency was validated across all dimensions; Cronbach's alpha coefficients ranged between $.766$ and $.926$, while the Composite Reliability values ranged between $.772$ and $.926$, surpassing the recommended threshold of $.70$. Furthermore, convergent accuracy was proven as the Average Variance Extracted exceeded the recommended $.50$ floor across all seven constructs, ranging from $.505$ (Perceived Benefits) to $.689$ (Privacy).

\subsubsection{Discriminant Validity}
We verified discriminant validity by computing the Heterotrait-Monotrait (HTMT) correlation ratios. The HTMT metric provides a more stringent and structurally accurate assessment of construct distinctiveness than traditional criteria \cite{ab2017discriminant, voorhees2016discriminant}. 

As detailed in Table \ref{tab:htmt_matrix}, all latent construct pairs safely met discriminant requirements. Every calculated HTMT ratio fell below the conservative methodological threshold of $\le 0.85$, with values ranging from a minimum of 0.61 (Perceived Bias to Actual Use) to a peak of 0.82 (Trust to Privacy \& Security). While this establishes that each construct captures a unique behavioral dimension statistically, the proximity of the Trust and Privacy \& Security ratio to the 0.85 threshold highlights a conceptual closeness. It suggests that the boundary between these concepts may be somewhat porous in emotional-support contexts, indicating that continued theoretical attention is necessary to ensure they remain sufficiently discriminant in future research within this domain.

\begin{table*}[t]
\centering
\scriptsize
\setlength{\tabcolsep}{2pt}
\caption{Heterotrait-Monotrait (HTMT) Correlation Ratio Matrix}
\label{tab:htmt_matrix}
\begin{tabular}{lccccccc}
\toprule
\textbf{Construct} & \textbf{1 (TRU)} & \textbf{2 (PRI)} & \textbf{3 (PER)} & \textbf{4 (BIA)} & \textbf{5 (BEN)} & \textbf{6 (USE)} & \textbf{7 (HUM)} \\
\midrule
1. Trust (TRU) & -- & & & & & & \\
2. Privacy (PRI) & .82 & -- & & & & & \\
3. Personalization (PER) & .75 & .65 & -- & & & & \\
4. Perceived Bias (BIA) & .72 & .70 & .73 & -- & & & \\
5. Perceived Benefits (BEN) & .80 & .65 & .80 & .80 & -- & & \\
6. Actual System Use (USE) & .68 & .73 & .65 & .61 & .72 & -- & \\
7. Humanlikeness (HUM) & .75 & .73 & .73 & .72 & .78 & .75 & -- \\
\bottomrule
\end{tabular}
\end{table*}

\subsection{Structural Equation Modeling (SEM)}

To test the hypothesized relationships between the system attributes, mediating variables, and actual usage, we estimated a SEM using Maximum Likelihood estimation. We ran a full mediation model to test if system attributes influence actual use through the mediating variables (Trust and Perceived Benefits). We assessed the structural paths using the full user sample ($N = 1,343$), and we evaluated indirect effects using 2,000 bootstrap iterations to establish robust significance intervals.

\subsubsection{Structural Model Fit}
We evaluated the overall goodness-of-fit of the structural model using standard psychometric indices. While the normed chi-square ($\chi^2/df = 4.424$, $\chi^2 = 2557.051$, $df = 578$) exceeded the strict threshold for a ``Good'' fit ($\leq 3.0$), it remained well within acceptable parameters ($\leq 5.0$). Importantly, this was compensated for by strong absolute and incremental fit indices, which confirmed a robust structural configuration. The Comparative Fit Index (CFI) of $.934$ and the Tucker-Lewis Index (TLI) of $.928$ were both above the $.90$ benchmark. Furthermore, the Root Mean Square Error of Approximation (RMSEA) of $.051$ (90\% CI: [$.049$, $.053$]) indicated an acceptable overall model fit, mitigating concerns regarding the elevated normed chi-square.

\subsubsection{Hypothesis Testing and Path Analysis}
The structural path estimates (summarized in Table \ref{tab:sem_results}) provide significant insights into the fully mediated drivers of LLM adoption for emotional support.

\begin{table*}[t]
\centering
\small
\caption{Structural Model Path Analysis: Direct and Indirect Effects}
\label{tab:sem_results}
\resizebox{\textwidth}{!}{%
\begin{tabular}{@{}lllccccc@{}}
\toprule
\textbf{Hyp.} & \multicolumn{2}{l}{\textbf{Direct Structural Paths}} & \textbf{Std. $\beta$} & \textbf{Estimate (B)} & \textbf{S.E.} & \textbf{C.R. ($t$-value)} & \textbf{Outcome} \\
\midrule
\multicolumn{8}{@{}l}{\textbf{Paths to Actual System Use}} \\
H1  & Perceived Benefits & $\rightarrow$ Actual System Use & .476 & .766 & .072 & 10.668*** & Supported \\
H2  & Trust & $\rightarrow$ Actual System Use & .348 & .428 & .053 & 8.058*** & Supported \\
\midrule
\multicolumn{8}{@{}l}{\textbf{Paths to Perceived Benefits}} \\
H3a & Humanlikeness & $\rightarrow$ Perceived Benefits & .263 & .179 & .025 & 7.181*** & Supported \\
H4a & Personalization & $\rightarrow$ Perceived Benefits & .312 & .296 & .039 & 7.571*** & Supported \\
H5a & Privacy & $\rightarrow$ Perceived Benefits & .138 & .097 & .021 & 4.543*** & Supported \\
H6a & Perceived Bias & $\rightarrow$ Perceived Benefits & -.297 & -.251 & .032 & -7.873*** & Supported \\
\midrule
\multicolumn{8}{@{}l}{\textbf{Paths to Trust}} \\
H3b & Humanlikeness & $\rightarrow$ Trust & .153 & .136 & .032 & 4.254*** & Supported \\
H4b & Personalization & $\rightarrow$ Trust & .261 & .325 & .050 & 6.486*** & Supported \\
H5b & Privacy & $\rightarrow$ Trust & .486 & .445 & .030 & 14.689*** & Supported \\
H6b & Perceived Bias & $\rightarrow$ Trust & -.111 & -.123 & .040 & -3.064** & Supported \\
\midrule
\midrule
\multicolumn{3}{l}{\textbf{Indirect Structural Paths (Bootstrapped)}} & \textbf{Std. $\beta$ (Ind)} & \textbf{Lower BC} & \textbf{Upper BC} & \textbf{$p$-value} & \textbf{Outcome} \\
\midrule
-- & Humanlikeness & $\rightarrow$ Actual System Use & .178 & .113 & .245 & $p < .01$ & Supported \\
-- & Personalization & $\rightarrow$ Actual System Use & .239 & .164 & .333 & $p < .01$ & Supported \\
-- & Privacy & $\rightarrow$ Actual System Use & .235 & .159 & .311 & $p < .01$ & Supported \\
-- & Perceived Bias & $\rightarrow$ Actual System Use & -.180 & -.253 & -.102 & $p < .01$ & Supported \\
\bottomrule
\multicolumn{8}{@{}l@{}}{\footnotesize \textit{Note:} Std. $\beta$ = Standardized Regression Weight; Estimate (B) = Unstandardized Regression Weight;} \\
\multicolumn{8}{@{}l@{}}{\footnotesize S.E. = Standard Error; C.R. = Critical Ratio. Indirect effects generated via 2,000 bootstrap iterations; } \\
\multicolumn{8}{@{}l@{}}{\footnotesize Bias-Corrected (BC) 95\% Confidence Intervals (Lower/Upper bounds) represent standardized estimates. } \\
\multicolumn{8}{@{}l@{}}{\footnotesize Perceived Bias paths were hypothesized to be negative. ** $p < .01$, *** $p < .001$.}
\end{tabular}%
}
\end{table*}

Regarding the mediating variable of \textbf{Trust}, all four system attributes showed significant effects. Privacy emerged as the dominant positive driver of trust ($\beta = .486$, $p < .001$), followed by Personalization ($\beta = .261$, $p < .001$) and Humanlikeness ($\beta = .153$, $p < .001$). As hypothesized, Perceived Bias significantly damaged Trust ($\beta = -.111$, $p = .002$). 

Similarly, for \textbf{Perceived Benefits}, all system attributes yielded significant relationships. Personalization was the strongest positive predictor of perceived benefits ($\beta = .312$, $p < .001$), followed closely by Humanlikeness ($\beta = .263$, $p < .001$) and Privacy ($\beta = .138$, $p < .001$). Conversely, Perceived Bias significantly diminished the perception of system benefits ($\beta = -.297$, $p < .001$).

When examining the direct drivers of \textbf{Actual System Use}, the full mediation structure revealed a clear dual-pathway to adoption. Both Perceived Benefits ($\beta = .476$, $p < .001$) and Trust ($\beta = .348$, $p < .001$) acting as direct antecedents significantly drove actual usage.

\subsubsection{Mediation and Indirect Effects}
To confirm the mediating roles of Trust and Perceived Benefits, we analyzed the standardized indirect effects via bootstrapping. The analysis confirmed that all system attributes exert highly significant indirect effects on Actual System Use. Personalization ($\beta_{ind} = .239$, $p < .01$) and Privacy ($\beta_{ind} = .235$, $p < .01$) had the strongest indirect positive impacts on usage, followed by Humanlikeness ($\beta_{ind} = .178$, $p < .01$). Finally, Perceived Bias exhibited a significant negative indirect effect on system use ($\beta_{ind} = -.180$, $p < .01$), confirming that bias deters usage by simultaneously eroding user trust and obscuring perceived benefits.

\begin{figure}[t]
\centering
\includegraphics[width=0.7\textwidth]{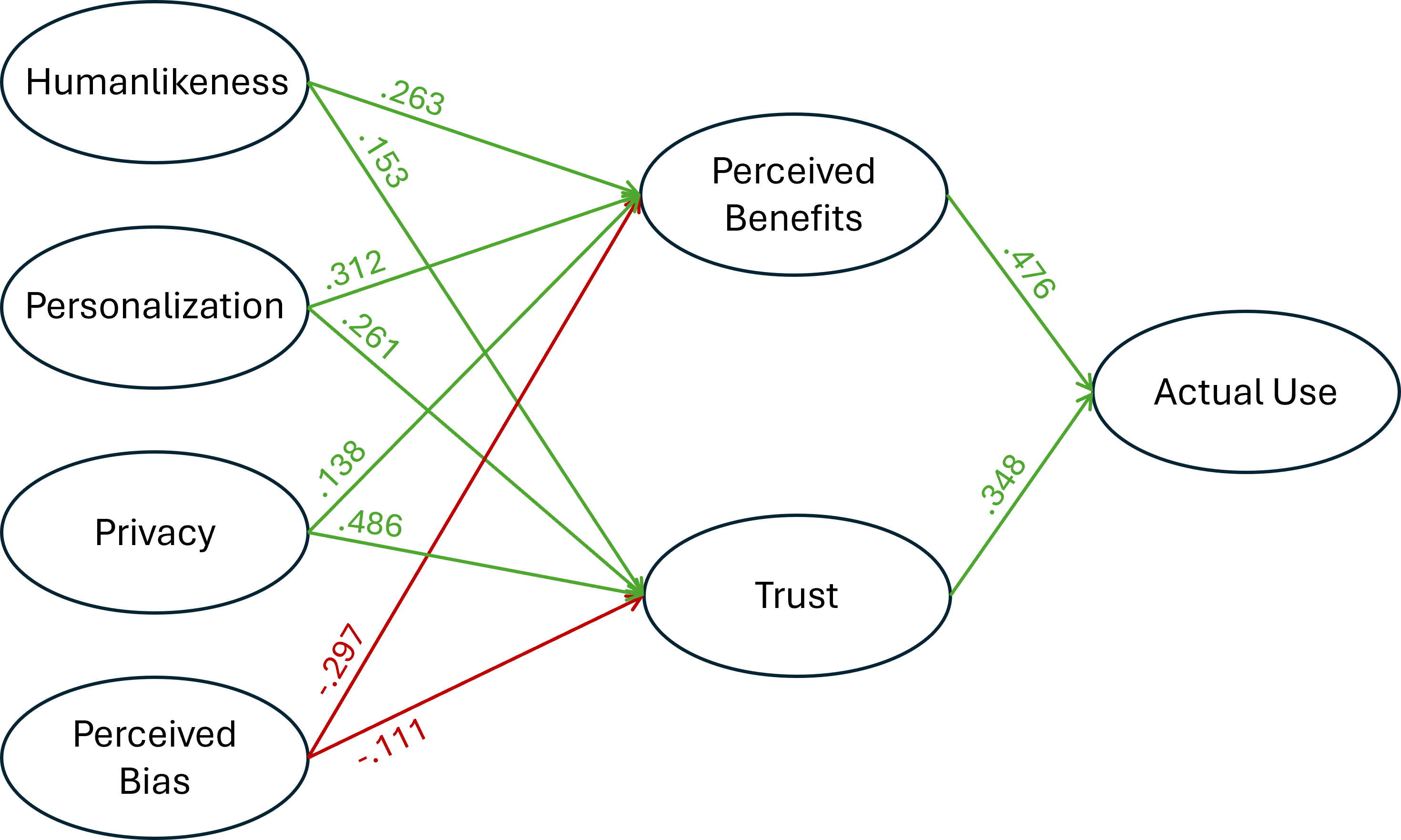}
\caption{The final Structural Equation Model. Note: All specified structural paths are significant at $p < .01$ or better. Standardized path coefficients ($\beta$) are displayed on the arrows. Positive effects in green, Negative effects in red.}
\label{fig:research_model_results}
\end{figure}

The complete path dynamics, including all significant direct relationships, are visually mapped in Figure \ref{fig:research_model_results}.

\subsection{Multi-Group Analysis (MGA)}

To understand how different groups of people perceive and use LLMs for emotional and mental wellbeing support, we conducted a series of multi-group analyses (MGA) using AMOS. This allowed us to see whether the relationships in our structural model depend on the users' background. We tested the model across five grouping variables ($N = 1,343$): gender (2 groups), age group (3 groups), cross-national region (3 groups), socioeconomic status (3 groups), and highest education level achieved (3 groups). Participants who selected "Prefer not to say" were excluded from those specific demographic analyses. Cross-national regions are based on the groups identified in Section~\ref{sec:sociodemographic_background}: the Anglosphere (UK, USA), Southern Europe (Italy, Spain), and Western/Central Europe (France, Germany, the Netherlands). Table \ref{tab:mga_sample_summary} summarizes the group configurations and sample distributions.

\begin{table}[t]
\centering
\caption{Summary of Sample Configurations for Multi-Group Analysis}
\label{tab:mga_sample_summary}
\resizebox{\textwidth}{!}{%
\begin{tabular}{llcc}
\toprule
\textbf{Grouping Variable} & \textbf{Sub-Group / Description} & \textbf{\textit{N}} & \textbf{\%} \\ \midrule
\multirow{2}{*}{Gender} & Male & 681 & 50.8\% \\
 & Female & 660 & 49.2\% \\ \midrule
\multirow{3}{*}{Age Group} & Early Adulthood (18–34 years) & 520 & 38.7\% \\
 & Middle Adulthood (35–44 years) & 370 & 27.5\% \\
 & Late Adulthood (45–65+ years) & 453 & 33.7\% \\ \midrule
\multirow{3}{*}{Cross-National Region} & Anglosphere (UK, USA) & 368 & 27.4\% \\
 & Southern Europe (Italy, Spain) & 398 & 29.6\% \\
 & Western/Central Europe (France, Germany, Netherlands) & 577 & 43.0\% \\ \midrule
\multirow{3}{*}{Socioeconomic Status (SES)} & Low SES (MacArthur Scale 1–4) & 190 & 14.1\% \\
 & Medium SES (MacArthur Scale 5–7) & 727 & 54.1\% \\
 & High SES (MacArthur Scale 8–10) & 426 & 31.7\% \\ \midrule
\multirow{3}{*}{Education Level} & No College Degree & 377 & 28.2\% \\
 & Undergraduate Degree & 446 & 33.3\% \\
 & Post-graduate \& Professional Degree & 516 & 38.5\% \\
\bottomrule
\multicolumn{4}{l}{\small \textit{Note:} Sub-group totals may vary slightly due to localized "Prefer not to say" exclusions.}
\end{tabular}%
}
\end{table}

\subsubsection{Measurement Invariance Testing}

Before comparing the structural relationships across different user groups, we evaluated both the overall model fit and the measurement invariance (specifically, metric invariance) for each configuration. This step ensures that the baseline multi-group models adequately represent the data and that our latent constructs (such as Trust, Humanlikeness, and Perceived Benefits) are interpreted consistently by all participants. 

As shown in Table \ref{tab:measurement_invariance}, all five baseline configurations demonstrated excellent overall model fit. The RMSEA values were comfortably below the standard 0.05 threshold, and the CFI values aligned closely to acceptable benchmarks (ranging from .888 to .927). 

To establish metric invariance, a chi-square difference test ($\Delta\chi^2$) was conducted comparing the unconstrained model against the constrained measurement weights model. A $p$-value greater than .05 indicates that factor loadings do not significantly differ between groups, confirming that the survey items measure the constructs identically. All five grouping variables successfully passed this test, providing a valid statistical foundation to evaluate cross-group structural variations.

\begin{table*}[t]
\centering
\small
\caption{Consolidated Model Fit and Measurement Invariance Results}
\label{tab:measurement_invariance}

\begin{tabular}{lcccc}
\toprule
& \multicolumn{2}{c}{\textbf{Baseline Model Fit}}
& \multicolumn{2}{c}{\textbf{Invariance Delta Test ($\Delta\chi^2$)}} \\
\cmidrule(r){2-3}
\cmidrule(l){4-5}

\textbf{Grouping Variable}
& \textbf{CFI}
& \textbf{RMSEA}
& \textbf{\textit{p}-value}
& \textbf{Invariance Supported?} \\
\midrule

Gender & .927 & .038 & .673 & Yes (Full metric) \\
Age Group & .909 & .034 & .354 & Yes (Full metric) \\
Cross-National Region & .888 & .035 & .294 & Yes (Full metric) \\
Socioeconomic Status (SES) & .908 & .033 & .623 & Yes (Full metric) \\
Education Level & .917 & .032 & .213 & Yes (Full metric) \\

\bottomrule
\multicolumn{5}{p{\linewidth}}{\small \textit{Note:} Invariance chi-square nested tests are evaluated directly against the unconstrained baseline model.}
\end{tabular}

\end{table*}

\subsubsection{Structural Path Comparisons}

The omnibus test for structural weights showed significant differences across the demographic configurations: gender ($p = .009$), age group ($p = .001$), cross-national region ($p = .000$), and education level ($p = .000$). Socioeconomic status demonstrated a marginal overall structural variance ($p = .080$); however, distinct pair-wise differences emerged at the item level. These results indicate that the factors driving user trust and perceived benefits, and how those mediators ultimately trigger actual usage, fluctuate depending on the user's background.

\begin{table}[t]
\centering
\caption{Consolidated Multi-Group Structural Path Comparisons (Significant Differences)}
\label{tab:structural_path_comparisons}
\resizebox{\textwidth}{!}{%
\begin{tabular}{llccccl}
\toprule
\textbf{Grouping} & \textbf{Structural Path} & \textbf{Est. G1} & \textbf{Est. G2} & \textbf{Est. G3} & \textbf{Result Summary} \\ \midrule
\textbf{Gender} & Privacy $\rightarrow$ Trust & .368 & .516 & -- & Stronger for Females (G2) \\
\textit{G1: Male} & Humanlikeness $\rightarrow$ Perceived Benefits & .114 & .232 & -- & Stronger for Females (G2) \\ 
\textit{G2: Female} & & & & & \\ \midrule
\textbf{Age Group} & Humanlikeness $\rightarrow$ Trust & .183 & .230 & .027 & Non-significant for Late Adulthood (G3) \\
\textit{G1: 18-34} & Privacy $\rightarrow$ Trust & .376 & .428 & .514 & Significantly stronger for Late Adulthood (G3) \\
\textit{G2: 35-44} & Trust $\rightarrow$ Actual System Use & .416 & .708 & .165 & Strongest for Middle (G2); Weakest for Late (G3) \\
\textit{G3: 45-65+} & Humanlikeness $\rightarrow$ Perceived Benefits & .113 & .169 & .235 & Significantly stronger for Late Adulthood (G3) \\
 & Personalization $\rightarrow$ Perceived Benefits & .333 & .462 & .148 & Drops significantly for Late Adulthood (G3) \\
 & Privacy $\rightarrow$ Perceived Benefits & .089 & .027 & .153 & Stronger for Late (G3) compared to Middle (G2) \\
 & Perceived Benefits $\rightarrow$ Actual System Use & .738 & .351 & 1.179 & Exceptionally strong for Late Adulthood (G3) \\ \midrule
\textbf{Region} & Humanlikeness $\rightarrow$ Trust & .303 & .044 & .139 & Strongest in Anglosphere (G1); N/S in G2 \\
\textit{G1: Anglo} & Personalization $\rightarrow$ Trust & .124 & .312 & .471 & Strongest in W/C Europe (G3); N/S in G1 \\
\textit{G2: S. Eur} & Privacy $\rightarrow$ Trust & .576 & .384 & .459 & Significantly stronger in Anglosphere (G1) \\
\textit{G3: W/C Eur} & Perceived Bias $\rightarrow$ Trust & -.037 & -.332 & .034 & Only significantly damages S. Europe (G2) \\
 & Trust $\rightarrow$ Actual System Use & .418 & .205 & .594 & Stronger in W/C Europe (G3) vs. S. Europe (G2) \\
 & Personalization $\rightarrow$ Perceived Benefits & .241 & .193 & .471 & Strongest in W/C Europe (G3) \\
 & Perceived Bias $\rightarrow$ Perceived Benefits & -.217 & -.335 & -.137 & Most damaging in S. Europe (G2) \\ 
 & Perceived Benefits $\rightarrow$ Actual System Use & .561 & .996 & .580 & Exceptionally strong in S. Europe (G2) \\ \midrule
\textbf{SES} & Privacy $\rightarrow$ Trust & .611 & .388 & .475 & Stronger for Low SES (G1) vs. Med SES (G2) \\
\textit{G1: Low} & Humanlikeness $\rightarrow$ Trust & -.048 & .153 & .148 & Non-significant only for Low SES (G1) \\
\textit{G2: Mid} & Trust $\rightarrow$ Actual System Use & .145 & .364 & .482 & Non-significant only for Low SES (G1) \\
\textit{G3: High} & & & & & \\ \midrule
\textbf{Education} & Personalization $\rightarrow$ Trust & .194 & .494 & .318 & Exceptionally strong for Undergrad Degree (G2) \\
\textit{G1: No Deg} & Privacy $\rightarrow$ Trust & .556 & .416 & .376 & Strongest for No College Degree (G1) \\
\textit{G2: Undgrad} & Trust $\rightarrow$ Actual System Use & .077 & .556 & .511 & Non-significant only for No College Degree (G1) \\
\textit{G3: Postgr} & Humanlikeness $\rightarrow$ Perceived Benefits & .235 & .213 & .108 & Stronger for No Degree (G1) vs. Postgrad (G3) \\
 & Privacy $\rightarrow$ Perceived Benefits & .159 & .110 & .057 & Stronger for No Degree (G1) vs. Postgrad (G3) \\
 & Perceived Benefits $\rightarrow$ Actual System Use & 1.219 & .531 & .698 & Exceptionally strong for No Degree (G1) \\
\bottomrule
\multicolumn{6}{l}{\small \textit{Note:} Listed paths exhibit significant cross-group structural variations ($|C.R.| > 1.96$, $p < .05$) between at least two subgroups.}
\end{tabular}%
}
\end{table}

To pinpoint where these differences occur, we examined the Critical Ratios (C.R.) for differences between parameters. A path varies significantly between two groups if the absolute corresponding C.R. is greater than $1.96$ ($p < .05$). Table \ref{tab:structural_path_comparisons} presents all the paths that showed significant group variations.

\paragraph{Gender Differences}
Two structural paths operate differently between men and women. While Privacy positively impacts Trust for both groups, the effect is significantly stronger for female users (C.R. = $2.44$). 
This suggests that the well-documented gender gap 
in AI trust \cite{stephany2026women} is not a generalized disposition but is specifically mediated by 
differential sensitivity to privacy risk: women appear to treat perceived data safety 
as a prerequisite for relational engagement with the system in a way that men do not. 

Additionally, Humanlikeness has a significantly stronger positive impact on the formation of Perceived Benefits for females compared to males (C.R. = $2.37$), suggesting that the warmth and 
authenticity of the interaction are more central to women's appraisal of practical 
value. Together, these findings point toward gendered trust formation mechanisms 
rather than a uniform gender effect, with implications for how privacy communication 
and interaction design should be calibrated.

\paragraph{Age Group Differences}
Seven paths vary across generations. For younger and 
middle-aged adults, Humanlikeness contributes meaningfully to Trust ($\beta = .183$ 
and $.230$ respectively), but this effect disappears for late adulthood 
users ($\beta = .027$, non-significant). Older adults instead rely more heavily on 
Privacy to build Trust ($\beta = .514$) and on Humanlikeness to 
perceive practical benefits ($\beta = .235$). This suggests that they value human-like 
interaction as a functional quality rather than as a trust signal. 
Age also alters how usage is triggered: the pathway from Trust to Actual System Use collapses for Late Adulthood users 
($\beta = .165$, non-significant), while the pathway from Perceived Benefits to 
Actual System Use strengthens significantly ($\beta = 1.179$). Older adults, in 
other words, do not require trust in order to engage: they adopt the technology on 
the basis of immediate practical value (availability, accessibility, non-judgmental 
support) bypassing the relational evaluation that younger users appear to require.


\paragraph{Cross-National Region Differences}
Eight paths vary across geographical regions. These cross-national findings align with the cultural profiles outlined in 
Section~\ref{sec:sociodemographic_background}.
Humanlikeness is a substantially 
stronger antecedent of Trust in the Anglosphere ($\beta = .303$) than in either 
Southern Europe ($\beta = .044$, non-significant) or Western/Central Europe 
($\beta = .139$), consistent with the Anglosphere's low uncertainty avoidance and 
strong self-expression values, a cultural profile associated with readily extending 
social responses to non-human agents \cite{srite2006role}. On the other hand, Western/Central European users rely strongly on Personalization to build Trust. 

The exceptionally strong 
Privacy path to Trust in the Anglosphere ($\beta = .576$) appears 
counterintuitive given its low uncertainty avoidance but may reflect a regulatory 
explanation: in the absence of a comprehensive data-protection baseline comparable 
to the GDPR, users must evaluate privacy sensitivity themselves, turning perceived 
privacy into a personal gatekeeper for trust \cite{bellman2004international}.


Perceived Bias significantly damages Trust only in Southern Europe ($\beta = -.332$) 
and exerts its strongest negative effect on Perceived Benefits there as well 
($\beta = -0.335$). Southern Europe's high uncertainty avoidance and tradition of 
relational trust may mean that a system perceived as biased violates fairness and 
loyalty norms, leading to the rejection of its use for personal matters 
\cite{putnam1993making, banfield1958moral}. That same heritage of low institutional 
trust may explain why Southern European adoption is driven almost entirely by 
Perceived Benefits ($\beta = .996$) while the Trust-to-Use path is the weakest 
across all regions ($\beta = .205$) \cite{hofstede2001cultures}. Western/Central 
Europe presents a contrasting profile: Trust and Perceived Benefits are built 
primarily through Personalization ($\beta = .471$ for both), and Trust converts 
into Use most strongly of all regions ($\beta = .594$), consistent with the 
cluster's high institutional trust and consensus-oriented culture 
\cite{inglehart2005modernization}.

\paragraph{Socioeconomic Status (SES) Differences}
The omnibus test for SES structural weights did not reach conventional significance 
($p = .080$), and the following path-level differences should therefore be 
interpreted as exploratory rather than confirmatory. Applying a Bonferroni 
correction for three comparisons ($\alpha = .0167$, $|C.R.| > 2.39$), only the 
Privacy $\rightarrow$ Trust difference survives as a robust finding; 
the remaining two differences are reported tentatively and warrant replication in 
future work.

With that caveat, the pattern of SES differences mirrors the one observed for age, but 
along a resource axis. Privacy drives Trust most strongly for Low SES 
users ($\beta = .611$), significantly more so than for Medium SES users ($\beta = .388$, 
$C.R. = -2.47$), suggesting that users with fewer resources may be more aware of 
their vulnerability when disclosing personal information. Tentatively, Humanlikeness 
appears to provide no significant trust value for Low SES users ($\beta = -.048$, 
$p = .590$), diverging from Mid and High SES users for whom it is a positive trust 
driver ($\beta = .153$ and $.148$ respectively, $C.R. = 2.04$). Similarly, Trust 
does not translate significantly into Actual System Use for Low SES users 
($\beta = .145$, $p = .252$), in contrast to Mid and High SES users ($\beta = .364$ 
and $.482$ respectively, $C.R. = 2.23$). Taken together — and pending replication 
— these patterns suggest that Low SES users may bypass the relational pathway 
entirely, reaching adoption through perceived practical benefits. This is  significant because these are users for whom AI emotional support 
may substitute for professional care they cannot access or afford, yet they appear 
to engage on the basis of utilitarian value rather than calibrated relational trust. This pattern raises equity concerns about the quality and safety of their 
engagement.

\paragraph{Education Level Differences}
Six structural relationships are altered by educational background. 

Education presents a divide in adoption logic that mirrors the 
age and SES findings. Users without a college degree bypass Trust entirely ($\beta = .077$, 
non-significant) and rely on Perceived Benefits to drive system use 
($\beta = 1.219$), the strongest such effect across all demographic groups. They 
are also significantly more sensitive to Privacy when forming Trust 
($\beta = .556$) and more reliant on Humanlikeness to perceive practical value 
($\beta = .235$) than their degree-holding counterparts. 
In contrast, undergraduate 
and postgraduate users require Trust as a precondition for engagement 
($\beta = .556$ and $.511$ respectively), with undergraduate users showing an 
exceptionally strong Personalization-to-Trust path ($\beta = .494$). The 
convergence of the no-degree and Low SES patterns is notable: in both cases, 
users who may have the fewest alternative support resources are those least likely 
to have developed a trust relationship with the system before engaging 
with it.

\paragraph{Synthesis}
Taken together, these findings reveal a consistent structural divide. Users with 
higher education and socioeconomic resources, and younger adults, follow a 
trust-mediated adoption logic in which relational safety is a prerequisite for 
engagement. Users with lower education, lower SES, and older adults converge on 
a benefits-driven adoption logic that bypasses trust. Cultural background shapes 
which system attributes build trust and how strongly, with Humanlikeness 
being particularly important in the Anglosphere, Personalization in Western/Central Europe, and 
bias sensitivity in Southern Europe. These findings demonstrate that 
technology acceptance structures are not universal \cite{im2011international} and 
that applying a one-size-fits-all model to the design and deployment of AI 
emotional support risks delivering inadequate experiences to the 
users who may need such support most.

\section{Discussion}

\subsection{Theoretical Implications}

This study makes several contributions to the theoretical understanding of trust and adoption in human-AI interaction, with particular relevance for AI for emotional support research.

\subsubsection{Psychometric instrument for emotional support AI}

A key contribution of this work is the development and validation of a 
seven-construct psychometric instrument for assessing user perceptions of LLMs in 
emotional support contexts (see Appendix \ref{tab:survey_items}). This scale is designed to capture 
the social-affective and bias-related dimensions that determine whether users feel safe 
enough to disclose personal emotions to an LLM. The instrument 
and its constructs (including the novel Perceived Bias scale) are 
designed to be adaptable to other affective computing and HCI contexts beyond 
emotional support.

\subsubsection{Extending technology acceptance theory to affective AI contexts}

Technology acceptance models have been widely applied to understand AI usage, often treating technology adoption as a largely uniform process in which perceived usefulness drives behavioral intention across user populations \cite{waluyo2025ai}. Our findings extend these frameworks in the context of emotional support AI in two main aspects. First, by  
reframing ``usefulness'' as Perceived Benefits: the practical and emotional value 
users derive from AI support interactions. Second, by establishing a dual-pathway architecture
in which a utilitarian route (Perceived Benefits) and a relational route 
(Trust) jointly and independently carry system attributes into actual use. This 
framing is more than a relabeling: it shows that users reach the same behavioral 
outcome through fundamentally different psychological routes. The MGA further shows 
that which pathway dominates is not fixed across users. Educated and higher-income 
users require trust before engaging, while older adults and lower-SES users bypass 
it entirely, relying on perceived practical benefits. This heterogeneity calls for 
sociodemographic moderation to be treated as a core rather than peripheral component 
of technology acceptance frameworks, particularly when applied in high-stakes or 
sensitive contexts. Recent work has similarly called for more context-sensitive 
conceptualizations of trust in AI \cite{wang2026validation}, and our results specify 
where and for whom that sensitivity matters most.

\subsubsection{The emergence of Humanlikeness as a unified construct}

Prior HCI research has treated cognitive empathy, affective empathy, and 
anthropomorphism as related but conceptually distinct constructs \cite{shang2026perception}. 
Our EFA finds that users do not experience them this way: all three load onto a 
single latent dimension, which we named Humanlikeness. We contribute to the literature 
by showing that these constructs (studied extensively as separate) can also be theorized and 
measured as one. In conversational AI contexts, users appear to form a holistic 
impression — the system either feels human-like or it does not — rather than 
decomposing their experience into separable facets. This aligns with qualitative 
evidence that users attend primarily to interactional features such as conversational 
flow and authenticity rather than to more abstract properties like intentionality or 
consciousness \cite{schimmelpfennig2025humanlike}. Scales that treat empathy and 
anthropomorphism as separate constructs may therefore be introducing distinctions 
that do not correspond to users' actual perceptual experience. The Humanlikeness 
scale validated in this study offers an integrated alternative for affective AI research.

\subsubsection{Perceived Bias as an active Trust- and Benefit-degrading construct}

HCI research on algorithmic fairness has often focused on objective performance 
metrics \cite{madaio2022assessing} or examined fairness perceptions in task-oriented contexts such as hiring \cite{de2024ai}
and content recommendation \cite{katsaros2024online}. We introduce and validate the 
first psychometric scale capturing users' subjective perception of bias specifically 
in LLM-based emotional support. Importantly, Perceived Bias does not only reduce 
positive system evaluations: it simultaneously and independently degrades both Trust 
and Perceived Benefits. This dual negative effect means that bias perceptions go beyond 
simply slowing trust formation, actively eroding the relational and utilitarian 
foundations of adoption at the same time \cite{afroogh2024trust}. While tested in 
the emotional support context, the Perceived Bias scale is designed to be adaptable: 
bias perceptions are a relevant psychological antecedent wherever users interact with 
AI systems capable of producing culturally or demographically unequal outputs, and 
we encourage its deployment and further validation across other HCI contexts.

\subsubsection{Sociodemographic heterogeneity in trust formation}

Across all five dimensions examined in the MGA, trust formation diverges in ways that are not 
arbitrary but theoretically interpretable. The stronger influence of Privacy on women's trust formation adds empirical specificity to the well-documented finding that women report lower trust in AI technologies \cite{stephany2026women}, suggesting that this gap is mediated by different sensitivity to privacy risk rather than a general distrust of technology. This finding points toward targeted 
design responses rather than generic trust-building interventions.

The cross-national finding that Anglosphere users respond more strongly to Humanlikeness than European users extends cultural frameworks such as Hofstede's \cite{hofstede2010cultures, hofstede2001cultures} to the specific domain of affective AI, empirically grounding theoretical predictions about how individualism and uncertainty avoidance shape responses to non-human social agents.

Taken together, 
these findings demonstrate that the dual-pathway architecture established in this 
study is not universal: the weights of the trust and benefit pathways vary 
significantly across gender, age, education, socioeconomic status, and cultural 
region. While recent scholars have begun noting the influence of individual demographic variables like gender or age on affective AI frameworks \cite{sun17exploring, mantello2023bosses}, a systematic framework evaluating how these diverse realities alter the underlying structural dynamics of technology acceptance in emotional support contexts has remained missing. Consequently, this study provides a meaningful contribution to HCI and technology acceptance research by offering a multi-dimensional, demographically granular model of trust formation.

\subsection{Practical Implications}

The findings of this study carry important implications for the design, deployment, and governance of AI emotional support systems. We outline them below for the following main audiences: designers, developers, and researchers.

For designers and developers, the structural model identifies three system attributes that reliably build user trust (Privacy, Personalization, and Humanlikeness) and one that degrades it (Perceived Bias). 
Privacy emerges as the most actionable starting point: it is under direct engineering control, and our MGA shows it to be especially important for women's trust formation. Concretely, this means sharing data handling practices transparently within the interaction interface itself, rather than relegating them to terms-of-service documents users often do not read. Furthermore, consent should be presented at the moment of emotional disclosure, not at onboarding. 

Regarding Personalization, the fact that users cite lack of memory across sessions as a key shortcoming \cite{chaudhry2024user} suggests that even lightweight continuity mechanisms (remembering a user's preferred coping strategies or previously disclosed stressors) could yield meaningful trust gains.

Humanlikeness, while a strong trust driver overall, should be calibrated with care: our MGA shows that Anglosphere users respond more strongly to humanlike cues than European users, suggesting that the optimal level of anthropomorphic expressiveness is not universal and should ideally be user-adjustable. 

Finally, Perceived Bias should be treated as a first-class design concern rather than a post-hoc audit item. This means building demographic sensitivity checks into response generation pipelines, acknowledging cultural differences and  pluralism in system responses, and avoiding single-viewpoint framings of wellbeing that implicitly privilege particular cultural norms.

For researchers, the instrument we developed and validated (particularly the Perceived Bias scale and the emergent Humanlikeness construct) provides a reusable psychometric foundation for future HCI studies in affective AI contexts. We encourage researchers to deploy and further validate this instrument across non-Western populations, as discussed in the limitations. 
Beyond replication, our MGA finding that older adults and lower-SES users reach adoption through perceived practical benefits while bypassing trust entirely has design research implications: it suggests that these groups may be particularly exposed to harm, since they engage with systems they have not yet developed a calibrated trust relationship with. Research that examines how to foster appropriate trust in these populations is especially needed.

\subsection{Limitations and Future Work}

This study advances the measurement and structural understanding of trust in AI emotional support, but several limitations should be acknowledged when interpreting the findings.
A first concern relates to self-report methodology. All constructs (including the outcome variable, Actual System Use) were measured through self-report Likert-scale items. This introduces the risk of social desirability bias, whereby respondents may overreport engagement with AI emotional support due to its perceived normalcy, or underreport it due to stigma around seeking mental health assistance through non-human agents. Future research should triangulate self-report data with behavioral measures, such as log-based usage frequencies.

A second limitation concerns the geographic scope of the sample. Although our seven-country design represents a methodological advance over the single-country convenience samples that often dominate the literature, all countries belong to Western, educated, industrialized, rich, and democratic (WEIRD) contexts. The cultural contrasts we identify (for instance, between Anglosphere and European users in their responsiveness to Humanlikeness cues) are therefore bounded within a relatively narrow band of the global cultural spectrum. Populations in East Asia, Sub-Saharan Africa, Latin America, and South Asia represent the majority of global LLM users yet remain absent from the evidence base. The cross-national patterns identified here, and particularly the MGA findings regarding Humanlikeness and Privacy, should therefore not be generalized beyond Western contexts. 

Third, the Perceived Bias construct introduced here is novel and requires further development. While it demonstrated good psychometric properties and a significant negative effect on Trust, the scale captures perceived rather than objective bias. Future research should examine the relationship between measured model bias (for example through systematic prompt auditing across demographic groups) and users' subjective perception of that bias, which may diverge substantially. Furthermore, the near-absence of gender-minority participants ($n=2$) precluded their inclusion in the gender multi-group analysis, representing a  sampling gap given that this construct encompasses assumptions about gender and sexual orientation.

\section{Conclusion}

This study addressed two gaps in HCI and AI for emotional 
support literature: the absence of validated psychometric instruments for assessing 
user perceptions of LLMs in affective support contexts, and the lack of large-scale, 
sociodemographically diverse evidence on how trust formation and adoption vary across 
user segments. Drawing on a sample of 1,343 users across seven countries, we 
developed and validated a seven-construct psychometric instrument, including a novel 
Perceived Bias scale, and tested a fully mediated structural model linking system 
attributes to Actual System Use through dual pathways of Trust and Perceived Benefits. 
The structural model confirmed that Privacy, Personalization, and 
Humanlikeness (an emergent unified construct integrating cognitive empathy, affective 
empathy, and anthropomorphism) drive Trust, while Perceived Bias simultaneously 
degrades both Trust and Perceived Benefits. A multi-group analysis across five 
demographic dimensions revealed that this architecture is not universal: trust-mediated 
adoption characterizes educated, higher-income, and younger users, while older adults 
and lower-socioeconomic groups bypass trust entirely, reaching engagement through 
perceived practical benefits alone. The user's background further shapes which system 
attributes build trust and how strongly. Together, these findings extend technology 
acceptance theory to affective AI contexts, provide a reusable measurement instrument 
for future HCI research, and demonstrate that equitable design of AI emotional support 
requires attending systematically to who the user is.

\section*{Disclosure Statement}
No potential conflict of interest was reported by the authors.


\section*{Data Availability Statement}
The data that support the findings of this study are available on request from the corresponding author.

\section*{Ethics Approval Statement}
\label{app:ethics}
This study received ethical approval from the Economics and Business Ethics Committee at the University of Amsterdam (Approval No: EB-18813).

\section*{Author Contribution Statement}

\textbf{Natalia Amat-Lefort:} Conceptualization, Data curation, Formal analysis, Investigation, Methodology, Writing – original draft, Writing – review \& editing. 
\textbf{Mert Yazan:} Conceptualization, Writing – original draft, Writing – review \& editing. 
\textbf{Amanda Cercas Curry:} Conceptualization, Supervision. 
\textbf{Flor Miriam Plaza-del-Arco:} Conceptualization, Project administration, Supervision, Writing – review \& editing. 

All authors provided final approval of the version to be published, revising it critically for intellectual content, and agree to be accountable for all aspects of the work.

\bibliographystyle{tfq}
 \bibliography{interacttfqsample} 

\clearpage
\appendix
\section{Psychometric instrument for emotional support AI}\label{app:quest}

\begin{center}
\scriptsize
\setlength{\tabcolsep}{5pt}
\renewcommand{\arraystretch}{1.2}
\begin{longtable}{>{\raggedright\arraybackslash}p{2.2cm}
>{\raggedright\arraybackslash}p{9.0cm}
>{\raggedright\arraybackslash}p{1.0cm}
>{\raggedright\arraybackslash}p{2.0cm}}

\caption{Survey Constructs, Item Codes, Statements, and Sources.} \label{tab:survey_items} \\
\multicolumn{4}{p{14.2cm}}{\vspace{-2mm}\footnotesize \textit{Note:} Items formatted in \textcolor{gray}{\sout{gray strikethrough}} were removed during the Exploratory Factor Analysis (EFA) or Confirmatory Factor Analysis (CFA) purification processes to ensure structural integrity.\vspace{2mm}} \\
\toprule
\textbf{Construct} & \textbf{Questionnaire Item} & \textbf{Code} & \textbf{Source} \\
\midrule
\endfirsthead

\caption*{(Continued) Survey Constructs, Item Codes, Statements, and Sources.} \\
\toprule
\textbf{Construct (Continued)} & \textbf{Questionnaire Item} & \textbf{Code} & \textbf{Source} \\
\midrule
\endhead

\midrule
\multicolumn{4}{r}{\textit{Continued on next page}} \\
\endfoot
\bottomrule
\endlastfoot

\multicolumn{4}{l}{\textbf{Humanlikeness} \textit{(Merged from Anthropomorphism, Cognitive \& Affective Empathy)}} \\
& \textcolor{gray}{\sout{The chatbot feels friendly and approachable.}} & \textcolor{gray}{ANT1} & \textcolor{gray}{\cite{bartneck2009measurement}} \\
& The chatbot gives me the impression that it has intentions or feelings. & ANT2 & \cite{bartneck2009measurement} \\
& I sometimes feel like I’m talking to a person, not a program. & ANT3 & \cite{bartneck2009measurement} \\
& The chatbot feels more like a companion than a tool. & ANT4 & \cite{bartneck2009measurement} \\
& Overall, the chatbot seems human-like in how it interacts with me. & ANT5 & \cite{bartneck2009measurement} \\
& \textcolor{gray}{\sout{The chatbot clearly informs me that it is a program and not a human.}} & \textcolor{gray}{ANT6} & \textcolor{gray}{Authors} \\
& I feel the chatbot understands how I am feeling and what I am thinking. & COG1 & \cite{reniers2011qcae} \\
& \textcolor{gray}{\sout{The chatbot tries to see things from my perspective before giving advice.}} & \textcolor{gray}{COG2} & \textcolor{gray}{\cite{reniers2011qcae}} \\
& The chatbot takes my feelings into account before it replies. & COG3 & \cite{reniers2011qcae} \\
& The chatbot supported me in coping with an emotional situation. & COG4 & \cite{schmidmaier2024} \\
& The chatbot expresses emotions that match the mood of our conversation (e.g., cheerful when appropriate, serious when needed). & AFF1 & \cite{reniers2011qcae} \\
& The chatbot shows emotional involvement when I share my problems. & AFF2 & \cite{reniers2011qcae} \\
& The chatbot expresses sadness or empathy appropriately when I share emotional experiences. & AFF3 & \cite{reniers2011qcae} \\
& Overall, the chatbot communicates in a way that shows empathy. & AFF4 & \cite{reniers2011qcae} \\
\midrule

\multicolumn{4}{l}{\textbf{Privacy}} \\
& \textcolor{gray}{\sout{The chatbot clearly communicates its limitations in providing emotional support.}} & \textcolor{gray}{PRI1} & \textcolor{gray}{Authors} \\
& I trust the chatbot to handle sensitive information safely. & PRI2 & \cite{marimon2024} \\
& I feel confident that my interactions with the chatbot are private. & PRI3 & \cite{marimon2024} \\
& Overall, I believe this chatbot is safe and respectful of my privacy. & PRI4 & Authors \\
& I feel safe sharing my emotions with the chatbot. & PRI5 & \cite{marimon2024} \\
& \textcolor{gray}{\sout{The chatbot provides advice on when to seek professional assistance or personal support from others.}} & \textcolor{gray}{PRI6} & \textcolor{gray}{Authors} \\
\midrule

\multicolumn{4}{l}{\textbf{Personalization}} \\
& \textcolor{gray}{\sout{The chatbot understands my goals.}} & \textcolor{gray}{PER1} & \textcolor{gray}{\cite{schmidmaier2024}} \\
& The chatbot adapts its responses to my personal preferences over time. & PER2 & \cite{marimon2024} \\
& The chatbot remembers relevant details from past interactions. & PER3 & \cite{marimon2024} \\
& The chatbot customizes its suggestions based on my needs. & PER4 & \cite{marimon2024}, \cite{schmidmaier2024} \\
\midrule

\multicolumn{4}{l}{\textbf{Perceived Bias}} \\
& The chatbot acknowledges and respects my perspectives or viewpoints. & BIA1 & Authors \\
& The chatbot avoids making biased assumptions about my religious or cultural background. & BIA2 & Authors \\
& The chatbot avoids making biased assumptions about my gender. & BIA3 & Authors \\
& The chatbot avoids making biased assumptions about my sexual orientation. & BIA4 & Authors \\
& \textcolor{gray}{\sout{The chatbot clearly indicates when it may have biases (e.g., cultural or contextual gaps) in its knowledge.}} & \textcolor{gray}{BIA5} & \textcolor{gray}{Authors} \\
& \textcolor{gray}{\sout{The chatbot acknowledges that there are multiple valid perspectives on well-being and doesn’t push a single viewpoint.}} & \textcolor{gray}{BIA6} & \textcolor{gray}{Authors} \\
& \textcolor{gray}{\sout{Overall, the chatbot's responses seem impartial and unbiased.}} & \textcolor{gray}{BIA7} & \textcolor{gray}{Authors} \\
& The chatbot understands my cultural context when responding to my concerns. & BIA8 & Authors \\
\midrule

\multicolumn{4}{l}{\textbf{Perceived Benefits}} \\
& \textcolor{gray}{\sout{The chatbot helps me feel more confident in managing emotional difficulties.}} & \textcolor{gray}{BEN1\_1} & \textcolor{gray}{Authors} \\
& \textcolor{gray}{\sout{The chatbot challenges my thinking in a constructive way.}} & \textcolor{gray}{BEN1\_2} & \textcolor{gray}{Authors} \\
& The chatbot helps me consider alternative viewpoints on my concerns. & BEN1\_3 & Authors \\
& The chatbot provides recommendations or advice that are useful for my specific situations. & BEN1\_4 & Authors \\
& The chatbot helps me identify actionable steps I can take to improve my mental well-being. & BEN1\_5 & Authors \\
& The chatbot's availability (e.g., 24/7 access) is a significant benefit for my well-being support. & BEN2\_1 & Authors \\
& I feel that I can express my feelings openly without fear of judgment. & BEN2\_2 & Authors \\
& \textcolor{gray}{\sout{Using this chatbot has improved my ability to manage my well-being concerns.}} & \textcolor{gray}{BEN2\_3} & \textcolor{gray}{Authors} \\
& \textcolor{gray}{\sout{The fact that this chatbot is free (or less expensive than traditional therapy) is a major benefit for me.}} & \textcolor{gray}{BEN2\_4} & \textcolor{gray}{Authors} \\
& Overall, I find the chatbot to be a valuable tool for supporting my mental well-being. & BEN2\_5 & Authors \\
& \textcolor{gray}{\sout{After using the chatbot, I feel calmer or more relaxed.}} & \textcolor{gray}{BEN2\_6} & \textcolor{gray}{Authors} \\
\midrule

\multicolumn{4}{l}{\textbf{Trust}} \\
& I trust the information or support this chatbot provides. & TRU1 & \cite{marimon2024} \\
& The chatbot provides credible information. & TRU2 & \cite{marimon2024} \\
& I trust that this chatbot is honest and transparent about what it can and cannot do. & TRU3 & Authors \\
& Overall, I trust the chatbot. & TRU5 & \cite{schmidmaier2024} \\
\midrule

\multicolumn{4}{l}{\textbf{Actual System Use}} \\
& I use the chatbot frequently for wellbeing/emotional support. & USE1 & \cite{davis1989} \\
& My interactions with the chatbot for mental wellbeing or emotional support are a regular part of my routine. & USE2 & \cite{davis1989} \\
& When I need mental wellbeing or emotional support, the chatbot is usually one of the first resources I turn to. & USE3 & \cite{davis1989} \\

\end{longtable}
\end{center}

\section{Exploratory Factor Analysis Results} \label{app:EFA_tables}

\begin{table*}[htbp]
\centering
\scriptsize
\setlength{\tabcolsep}{6pt}
\renewcommand{\arraystretch}{1.2}
\caption{Rotated Component Matrix for Independent Variables (First Split Sample, $n = 671$)}
\label{tab:efa_independent}
\begin{tabular}{p{2.2cm} >{\centering\arraybackslash}p{1.8cm} >{\centering\arraybackslash}p{1.8cm} >{\centering\arraybackslash}p{1.8cm} >{\centering\arraybackslash}p{1.8cm}}
\toprule
 & \multicolumn{4}{c}{\textbf{Component}} \\ \cmidrule{2-5}
\textbf{Item Code} & \textbf{1 (HUM)} & \textbf{2 (BIA)} & \textbf{3 (PRI)} & \textbf{4 (PER)} \\
\midrule
AFF2 & \cellcolor{rowgray}\textbf{.742} & .199 & .100 & .183 \\
ANT2 & \cellcolor{rowgray}\textbf{.705} & .081 & .316 & .080 \\
ANT5 & \cellcolor{rowgray}\textbf{.702} & .272 & .264 & .100 \\
AFF3 & \cellcolor{rowgray}\textbf{.700} & .155 & .053 & .201 \\
COG1 & \cellcolor{rowgray}\textbf{.679} & .211 & .220 & .296 \\
AFF1 & \cellcolor{rowgray}\textbf{.674} & .220 & .092 & .267 \\
ANT3 & \cellcolor{rowgray}\textbf{.667} & .234 & .316 & .034 \\
AFF4 & \cellcolor{rowgray}\textbf{.635} & .336 & .157 & .240 \\
ANT4 & \cellcolor{rowgray}\textbf{.627} & .180 & .378 & .155 \\
COG4 & \cellcolor{rowgray}\textbf{.590} & .197 & .238 & .292 \\
COG3 & \cellcolor{rowgray}\textbf{.586} & .209 & .227 & .363 \\
COG2 & .519 & .246 & .166 & .436 \\
ANT1 & .453 & .332 & .181 & .307 \\ \midrule
BIA4 & .199 & \cellcolor{rowgray}\textbf{.763} & .078 & .135 \\
BIA2 & .176 & \cellcolor{rowgray}\textbf{.741} & .132 & .189 \\
BIA3 & .181 & \cellcolor{rowgray}\textbf{.726} & .169 & .133 \\
BIA7\textsuperscript{a} & .227 & \textbf{.644} & .191 & .112 \\
BIA1 & .325 & \cellcolor{rowgray}\textbf{.567} & .206 & .298 \\
BIA8 & .344 & \cellcolor{rowgray}\textbf{.558} & .225 & .237 \\
BIA6 & .193 & .532 & .180 & .364 \\
BIA5 & .195 & .498 & .249 & .248 \\ \midrule
PRI4 & .268 & .238 & \cellcolor{rowgray}\textbf{.787} & .203 \\
PRI2 & .248 & .204 & \cellcolor{rowgray}\textbf{.771} & .200 \\
PRI3 & .281 & .204 & \cellcolor{rowgray}\textbf{.762} & .205 \\
PRI5 & .371 & .235 & \cellcolor{rowgray}\textbf{.649} & .233 \\
PRI6 & .259 & .324 & .454 & .345 \\ \midrule
PER2 & .296 & .193 & .122 & \cellcolor{rowgray}\textbf{.698} \\
PER3 & .256 & .163 & .189 & \cellcolor{rowgray}\textbf{.660} \\
PER4 & .236 & .295 & .141 & \cellcolor{rowgray}\textbf{.660} \\
PER1 & .378 & .206 & .286 & .514 \\ \midrule
PRI1 & .157 & .202 & .395 & .459 \\
ANT6 & .057 & .294 & .257 & .412 \\
\bottomrule
\multicolumn{5}{p{10.2cm}}{\small \textit{Note:} Shaded cells represent items retained for the CFA phase. HUM = Humanlikeness (merged from ANT = Anthropomorphism, COG = Cognitive Empathy, AFF = Affect Empathy), BIA = Perceived Bias, PRI = Privacy, PER = Personalization.} \\
\multicolumn{5}{p{10.2cm}}{\small \textsuperscript{a}Item dropped post-EFA during CFA to improve discriminant validity.}
\end{tabular}
\end{table*}

\begin{table*}[t]
\centering
\scriptsize
\setlength{\tabcolsep}{6pt}
\renewcommand{\arraystretch}{1.2}
\caption{Rotated Component Matrix for Mediating and Response Variables (First Split Sample, $n = 671$)}
\label{tab:efa_response}
\begin{tabular}{p{2.2cm} >{\centering\arraybackslash}p{2.4cm} >{\centering\arraybackslash}p{2.4cm} >{\centering\arraybackslash}p{2.4cm}}
\toprule
 & \multicolumn{3}{c}{\textbf{Component}} \\ \cmidrule{2-4}
\textbf{Item Code} & \textbf{1 (BEN)} & \textbf{2 (USE)} & \textbf{3 (TRU)} \\
\midrule
BEN2\_1 & \cellcolor{rowgray}\textbf{.756} & .231 & .082 \\
BEN2\_4\textsuperscript{a} & \textbf{.691} & .176 & .185 \\
BEN1\_4 & \cellcolor{rowgray}\textbf{.670} & .205 & .327 \\
BEN2\_2 & \cellcolor{rowgray}\textbf{.661} & .241 & .235 \\
BEN1\_5 & \cellcolor{rowgray}\textbf{.617} & .217 & .367 \\
BEN1\_3 & \cellcolor{rowgray}\textbf{.585} & .165 & .398 \\
BEN2\_5 & \cellcolor{rowgray}\textbf{.577} & .457 & .320 \\
BEN2\_3 & .518 & .474 & .332 \\
BEN1\_1 & .461 & .438 & .370 \\ \midrule
USE2 & .180 & \cellcolor{rowgray}\textbf{.839} & .227 \\
USE3 & .243 & \cellcolor{rowgray}\textbf{.810} & .186 \\
USE1 & .257 & \cellcolor{rowgray}\textbf{.788} & .240 \\
BEN2\_6 & .474 & .521 & .314 \\
BEN1\_2 & .329 & .434 & .349 \\ \midrule
TRU1 & .244 & .277 & \cellcolor{rowgray}\textbf{.766} \\
TRU2 & .283 & .171 & \cellcolor{rowgray}\textbf{.757} \\
TRU3 & .248 & .222 & \cellcolor{rowgray}\textbf{.754} \\
TRU5 & .284 & .345 & \cellcolor{rowgray}\textbf{.699} \\
\bottomrule
\multicolumn{4}{p{10.2cm}}{\small \textit{Note:}  Shaded cells represent items retained for the CFA phase. BEN = Perceived Benefits, USE = Actual System Use, TRU = Trust.} \\
\multicolumn{4}{p{10.2cm}}{\small \textsuperscript{a} Item dropped post-EFA during CFA to improve discriminant validity.}
\end{tabular}
\end{table*}

\end{document}